\documentstyle[12pt, amsfonts, amssymb]{article}

\date{\today}
\begin{document}

\def\be{\begin{equation}}
\def\ee{\end{equation}}
\def\ba{\begin{eqnarray}}
\def\ea{\end{eqnarray}}
\def\lb{\label}

\def\a{\alpha}
\def\b{\beta}
\def\g{\gamma}
\def\d{\delta}
\def\s{\sigma}
\def\e{\varepsilon}
\def\l{\lambda}
\def\v{\varepsilon}
\def\L{\Lambda}
\def\T{\Theta}

\def\E{{\cal E}}

\def\C{\Bbb C}
\def\N{\Bbb N}
\def\R{\Bbb R}
\def\Z{\Bbb Z}

\def\bq{\overline{q}}

\def\id{\mbox{\rm 1\hspace{-3.5pt}I}}

\hyphenation{re-pre-sen-ta-ti-on}
\hyphenation{in-de-com-po-sa-ble}


\title{Indecomposable $U_q(sl_n)\,$ modules for $q^h =-1\,$ and BRS
intertwiners}

\author{P. Furlan\footnote{E-mail: furlan@trieste.infn.it}
\ $^{1,2}$\,,\,
L.K. Hadjiivanov\footnote{E-mail: lhadji@inrne.bas.bg}
\ $^{1,2,4}$\, and\,
I.T. Todorov\footnote{E-mail: todorov@inrne.bas.bg}
\ $^{3,4}$
\\
{\footnotesize
$^1$ Dipartimento di Fisica Teorica dell' Universit\`a di
Trieste, I-34100 Trieste, Italy}\\
{\footnotesize
$^2$ Istituto Nazionale di
Fisica Nucleare (INFN), Sezione di Trieste, Trieste, Italy}\\
{\footnotesize
$^3$ Laboratorio Interdisciplinare per le Scienze Naturali e
Umanistiche,}\\
{\footnotesize
International School for Advanced Studies, SISSA/ISAS, I-34014 Trieste,
Italy}\\
{\footnotesize
$^4$ Division of Theoretical Physics, Institute for Nuclear
Research and Nuclear Energy,}\\
{\footnotesize
Bulgarian Academy of Sciences, Tsarigradsko Chaussee
72, BG-1784 Sofia, Bulgaria
\footnote{Permanent address}}
}

\maketitle

\begin{abstract}

A class of indecomposable representations of
$U_q(s\ell_n)\,$ is considered for
$q\,$ an even root of unity ($q^h = -1\,$) exhibiting a similar
structure as (height $h\,$) indecomposable
lowest weight Kac-Moody modules associated with a chiral conformal field
theory. In particular, $U_q(s\ell_n)\,$ counterparts of the
Bernard-Felder BRS
operators are constructed for $n= 2, 3\,.$
For $n=2\,$ a pair of dual $d_2(h) = h\,$ dimensional  $U_q(s\ell_2)\,$
modules gives rise to a $2h\,$ dimensional indecomposable representation
including those studied earlier in the context of tensor product
expansions of irreducible representations.
For $n=3\,$ the interplay between the Poincar\'e-Birkhoff-Witt and
(Lusztig) canonical bases is exploited in the study of
$d_3(h) = \frac{h(h+1)(2h+1)}{6}\,$ dimensional indecomposable
modules and of the corresponding intertwiners.

\end{abstract}

\newpage


\textwidth = 16.5truecm
\textheight = 22.5truecm
\hoffset = -1truecm
\voffset = -2truecm

\section{Introduction}
\setcounter{equation}{0}
\renewcommand\theequation{\thesection.\arabic{equation}}

\medskip

Quantum group representations at $q\,$ a root of unity arise in
the study of chiral components of a Wess-Zumino-Novikov-Witten
(WZNW) \cite{W, KZ} current algebra model \cite{B, F1, F2, G, FG, FHT1,
FHT2, FHT3, FHIOPT}.
Such chiral models necessarily involve an extended phase space
with unphysical degrees of freedom. The Verma modules of a
Kac-Moody current algebra (or, rather, their Wakimoto extensions \cite{Wa})
were shown to give rise to a Becchi-Rouet-Stora (BRS) cohomology
\cite{BF, BMP}. A quantum group counterpart of this
construction was worked out for $U_q(s\ell_2)\,$ in \cite{FG} where
a complex of partially equivalent $U_q(s\ell_2)\,$ modules and
intertwining ``BRS" maps were exhibited. Motivated by this work
we study dual pairs of finite dimensional indecomposable
representations of $U_q(s\ell_n)\,$ and the intertwining maps
between them.

After some preliminaries about dual representations and linear and
antilinear antiinvolutions in $U_q(s\ell_n)\,$ (Section 2) we give in Section
3 a comprehensive study of
dual pairs of $h\,$ dimensional indecomposable $U_q(s\ell_2)\,$ modules for
\be
\lb{qh}
q^h=-1\,,\ q+\bq = 2\cos\frac{\pi}{h}\,.
\ee
Two inequivalent families of pairs,
$({\cal C}_p ,\,{\cal C}_{2h-p} )\,$ and
$({\cal V}_p ,\,{{\cal V}}_{2h-p} )\,$ are
considered. The modules ${\cal C}_p\,$ and ${\cal C}_{2h-p}\,$
are cyclic, ${\cal V}_p\,$ has one highest weight (HW) and two lowest
weight (LW) vectors for $0<p<h\,,$ while ${{\cal V}}_{2h-p}\,$ has
one LW and two HW vectors (the modules ${\cal C}_h\,$ and ${\cal V}_h\,$
being equivalent and irreducible). For both pairs we define intertwining
maps (``BRS operators")
$Q^{h-p} \,:\ {\cal C}_p\,\to\,{\cal C}_{2h-p}\quad (\,{\cal V}_p\,\to\,
{{\cal V}}_{2h-p}\,)\,$ whose $(h-p)$-dimensional kernels ${\cal
I}_{h-p}\,$ carry isomorphic irreducible representations (IR) of
$U_q(s\ell_2)\,.$ The ``physical" $p$-dimensional IR appears as a quotient
${\cal C}_p / {\cal I}_{h-p}\,\simeq {\cal V}_p / {\cal I}_{h-p}\,.$
Both ${\cal C}_p\,$ and ${\cal V}_p\,$ admit a unique invariant bilinear
form which gives rise to a non-degenerate inner product on the factor
space (Proposition 3.3).

In Section 4 we construct $2h$-dimensional indecomposable representations
${\cal D}_p\,$ and ${{\cal W}}_p\,$ of $U_q(s\ell_2)\,$
such that
${\cal C}_{2h-p}\,$ and ${{\cal V}}_{2h-p}\,$ appear as
$U_q(s\ell_2)\,$
invariant submodules while
their duals, ${\cal C}_p\,$ and ${\cal V}_p\,,$
are isomorphic to the quotient spaces
${\cal D}_p / {\cal C}_{2h-p}\,$ and
${{\cal W}}_p / {{\cal V}}_{2h-p}\,,$ respectively.
${{\cal W}}_p\,$ coincide with the
indecomposable $2h$-dimensional $U_q(s\ell_2)\,$ modules considered earlier
\cite{PS90, ReshTur, FK}.

In Section 5 after some general remarks about $U_q(s\ell_n)\,$ we
study the case of $n=3\,.$ We only consider the analogs of the pairs
$({\cal V}_p ,\,{{\cal V}}_{2h-p} )\,$ in this case constructing
$d_3(h)\,$ dimensional indecomposable $U_q(s\ell_3)\,$ modules ${\cal V}_{\bf
p}\ ({\bf p}=(p_{12} , p_{23} ),\ p_{i\, i+1}\in \N\,)\,$ where
\be
\lb{d3}
d_3(h) = \frac{h(h+1)(2h+1)}{6}\,.
\ee
The dual modules, ${{\cal V}}_{w_L{\bf p}}\,,$ are defined in terms
of the
longest element, $w_L = w_1 w_2 w_1\,,$ of the $s\ell_3\,$ Weyl group.
We give an explicit construction of the BRS operator
\be
\lb{BRSQ}
Q_{\bf p} :\
{\cal V}_{w_L{\bf p}}\ \rightarrow\ {\cal V}_{\bf p}
\ee
using the Poincar\'e-Birkhoff-Witt (PBW) basis in both modules. For
\be
\lb{p13}
(\, 1 <\, )\, p_{13} = p_{12} + p_{23}\, <\, h
\ee
$Q_{\bf p}\,$ maps the cosingular vector $| w_L{\bf p} ; 0, 0, 0 {\cal
i}\,\in {\cal V}_{w_L{\bf p}}\,$ to a singular (LW) vector in ${\cal
F}_{\bf p}\,$ which
belongs to the canonical (Lusztig-Kashiwara)
basis. Denote by ${\bf h}\,$ the weight $(p_{12} = h\,,\  p_{23} =
h\,)\,;$
the BRS property is expressed by the relation
\be
\lb{BRSpr}
Q_{{\bf h}+ w_L {\bf p}}\, Q_{\bf p}\, =\, 0\quad (
Q_{{\bf h}+ w_L {\bf p}}\,:\ {\cal V}_{\bf p}\ \rightarrow\
{\cal V}_{{\bf h}+ w_L {\bf p}}\, ).
\ee
The BRS cohomology is found to be trivial (for both $n=2\,$ and $n=3\,$):
\be
\lb{cohtriv}
{\rm Ker}\, Q_{{\bf h}+ w_L {\bf p}}\, =\, {\rm Im}\, Q_{\bf p}\ (\subset
{\cal V}_{\bf p}\, )\,,
\ee
${\rm Im}\, Q_{\bf p}\,$ and
${\rm Ker}\, Q_{{\bf h}+ w_L {\bf p}}\,$ defining
an invariant subspace of ${\cal V}_{\bf p}\,.$
This invariant subspace is shown to lie in the kernel of the invariant
hermitean form on ${\cal V}_{\bf p}\,.$


\vspace{5mm}

\section{Preliminaries. Dual representations and (co)singular vectors}

\setcounter{equation}{0}
\renewcommand\theequation{\thesection.\arabic{equation}}

\medskip

We shall assume, for definiteness, throughout this paper that
$q=e^{-i{{\pi}\over h}}\,$ with $h\,$ an integer greater than $n\,.$
Such $q\,$ appears naturally as quantum group deformation parameter
corresponding to the left chiral WZNW field in the conventions of
\cite{FHIOPT}. We shall use the notations
$\bq := q^{-1}\,, \quad [m] := {{q^m - \bq^m}\over {q - \bq}}\,$ and
$(m)_\pm := {{q^{\pm 2m} - 1}\over {q^{\pm 2} - 1}}
\equiv q^{\pm (m-1)} [m]\,$.

We start by fixing our conventions: $U_q(s\ell_n)\,$ is a Hopf algebra with
$4(n-1)\,$ (Chevalley)
generators $E_i\,,\,F_i\,,\, q^{H_i}\,, \,
q^{-H_i}\equiv \bq^{H_i}\,,\ i=1,2,\dots ,n-1\,$
satisfying
\be
\lb{Lambda}
q^{H_i} E_j = E_j q^{H_i + c_{ij}}\,,\quad
q^{H_i} F_j = F_j q^{H_i - c_{ij}}\,,\quad
q^{H_i} \bq^{H_i} = \id =
\bq^{H_i} q^{H_i}\,,
\ee
$c_{ij}\,$ being the $s\ell_n\,$ Cartan matrix,
\be
\lb{EFH}
[E_i\,,\, F_j ] = \d_{ij} [H_i]\,,
\ee
and the Serre relations
\ba
\lb{Serre}
&&E_{i+1\; i}  E_i = q E_i E_{i+1\; i}\quad{\rm for}\quad
E_{i+1\; i}:= E_i E_{i+1} - q E_{i+1} E_i\,,\ 1\le i\le n-2\,,\nonumber\\
{}\\
&& F_i F_{i\; i+1}  = q F_{i\; i+1} F_i\quad{\rm for}\quad
F_{i\; i+1}:= F_{i+1} F_i - q F_i F_{i+1}\,,\ 1\le i\le n-2\,.\nonumber
\ea

The {\em coproduct} $\Delta :\, U_q(s\ell_n) \rightarrow U_q(s\ell_n)\otimes
U_q(s\ell_n)\,$ and the {\em counit} $\e : U_q(s\ell_n) \rightarrow
\C\,$ are algebra homomorphisms fixed by their action on the generators:
\ba
\lb{Delta}
&&\Delta (q^{\pm H_i}) = q^{\pm H_i}\otimes q^{\pm H_i}\,,\nonumber\\
&&\Delta (E_i) = E_i\otimes q^{H_i} + \id \otimes E_i\,,\quad
\Delta (F_i) = F_i\otimes \id + {\bq}^{H_i}\otimes F_i\,;
\ea
\be
\lb{epsi}
\varepsilon (q^{\pm H_i}) = 1\,,\quad
\varepsilon (E_i) = 0 = \varepsilon (F_i)\,.
\ee
The {\em antipode} $\g : U_q(s\ell_n) \rightarrow U_q(s\ell_n)\,$ is an
algebra antihomomorphism (i.e. $\g (XY) = \g (Y) \g (X)\,)\,$
characterized by the property
\be
\lb{co}
\sum \gamma (X_1) X_2 = \sum X_1 \gamma (X_2) = \varepsilon (X)\, \id
\ \ \ {\rm for}\ \ \Delta (X) = \sum X_1 \otimes X_2\,;
\ee
with the above coproduct and counit this relation implies
\be
\lb{gamma}
\gamma (q^{\pm H_i}) = q^{\mp H_i}\,,\quad
\gamma (E_i) = - E_i\, \bq^{H_i}\,,\quad
\gamma (F_i) = - q^{H_i} F_i\,.
\ee
A convenient way to single out the two Borel (Hopf-)
subalgebras $U_q^F\,$ and $U_q^E\,$ of $U_q(s\ell_n)\,$ \cite{FRT}
(corresponding to the Gauss decomposition of the WZNW
monodromy matrix $M=M_+M_-^{-1}\,$ \cite{FHT3}) is
to arrange their generators in an
$n\times n\,$ matrix form where
\ba
\lb{mon}
&&\qquad M_\pm^{\pm 1}= N_\pm D\,,\qquad  D^i_j
=q^{d_i}\d_j^i\,,\nonumber\\
&&\sum_{j=1}^{n}\, d_j = 0 \,,\qquad
\sum_{j=1}^{i}\, d_j = - \sum_{j=1}^{n-1} (c^{-1})_{ij} H_j\,,\
\ i\le n-1\,,\nonumber\\
&&\qquad N_+= \left(\matrix{ 1 &(\bq -q)F_1 &(\bq -q)F_{12} &\ldots\cr
0 &1 &(\bq -q)F_2 &\ldots\cr
0 &0 &1 &\ldots\cr
\ldots &\ldots &\ldots &\ldots
\cr}\right)\,,\nonumber\\
&&\qquad N_-= \left(\matrix{ 1 &0 &0 &\ldots\cr
(\bq -q) E_1 &1 &0 &\ldots\cr
(\bq -q)E_{21} &(\bq -q)E_2 &1 &\ldots\cr
\ldots &\ldots &\ldots &\ldots\cr}\right)\,.
\ea
(In fact, the weight operators $q^{d_i}\,,$ together with
$E_i\,$ and $F_i\,,$ give rise to the so called {\em simply connected
\cite{DCK} quantum universal enveloping algebra} $U_q(s\ell_n)\,.$)
The subalgebra $U_q^F\,$ (resp., $U_q^E\,$) is generated by the
matrix elements of $M_+\,$ (resp., $M_-^{-1}\,$). The
coproduct of the latter
is expressed as the matrix multiplication of two copies of
$M_+\,$ (resp., $\, M_-\,$) while the antipode is related to the
inverse matrices:
\be
\lb{DeltaM}
\Delta ({M_\pm}^{\a~}_{~\b} ) = {M_\pm}^{\a~}_{~\sigma}\otimes
{M_\pm}^{\sigma ~}_{~\b}\,,\quad\   (M_-^{-1})^{\a~}_{~\b}\,
= \, \g ({M_-}^{\a~}_{~\b})\,.
\ee
We introduce a {\em grading} in $U_q(s\ell_n)\,$ (corresponding to
the {\em depth} of \cite{BF} ) ascribing degree $1\,$ to $E_i\,,\
0\,$ to $q^{\pm H_i}\,,\,$ and $-1\,$ to $F_i\,$ (so that, e.g.
$E_{21}\,$ and $F_{12}\,$ have degree $2\,$ and $-2\,,$
respectively); we denote by $U^-_q\ (\, U_q^+\,)\,$ the
subalgebra of $U^F_q\ (\, U_q^E\,)\,$ of elements of negative
(resp., positive) degree.

Next we introduce the concepts of (co)singular vectors and dual
indecomposable representations adapting to our case the
discussion in Section 4 of \cite{BF}.

Let ${\cal V}\,$ be a $U_q(s\ell_n)\,$ module. An eigenvector $v\in
{\cal V}\,$ of the Cartan generators $q^{H_i}\,$ (i.e. a {\em
weight vector}) is said to be {\em ($F$-) singular}
if it satisfies $U_q^-\, v \, =\, 0\,.$
It is called {\em cosingular} if there is no $v' \in {\cal V}\,$
such that $v\in U_q^+ v'\,.$
{\em Two cosingular vectors are equivalent} if their difference
belongs to $U_q^+ v_1\,$
for some $v_1\in {\cal V}\,$. The lowest
weight vector in a Verma module is both singular and cosingular.

These definitions are adapted for lowest weight
modules. For highest weight modules one can just reverse the
roles of $U_q^+ \,$ and $U_q^-\,.$

We shall be dealing in what follows with {\em indecomposable
$U_q(s\ell_n)\,$ modules} ${\cal V}_p\,$ corresponding -- in a way
that will be made clear below -- to integral dominant weights $\L\;
=\; p - \rho\,,$ and their duals. Here $p\,$ are the {\em shifted}
weights and $\rho \,$ is the half sum
of the positive roots; we shall use the (linearly dependent) ``barycentric"
coordinates of $p$
\be
\lb{p}
p = \{ (p_1, \dots \, ,p_n )\,;\, \sum_{i=1}^n  p_i = 0\,,\, p_{i\; i+1}
:= p_i - p_{i+1} \in {\N} \, = \{ 1, 2, \dots \} \ \}
\ee
fixed by the condition $\L \equiv
\sum_{i=1}^{n-1} \lambda_i \,\L^{(i)} =
\sum_{i=1}^{n-1} (p_{i\;i+1} - 1)\,\L^{(i)}\,,$
where $\L^{(i)}\,$ are the fundamental weights
$(\rho = \sum_{i=1}^{n-1} \, \Lambda^{(i)}\,)\,.$

There are several inequivalent
representations of this type which share the following properties.

\vspace{5mm}

\noindent
i)~${\cal V}_p\,$ is a direct sum of finite dimensional weight spaces,
\ba
\lb{wVp}
{\cal V}_p = {\oplus}_{\lambda \in \L_- + Q} {\cal V}_p^{(\lambda
)}\,,\qquad (q^{H_i} - q^{\lambda_i}) {\cal V}_p^{(\lambda )} = 0\quad
({\rm dim}\,{\cal V}_p^{(\lambda )} <\infty )
\ea
where $\L_- =\sum_{i=1}^{n-1}\, (1- p_{n-i\;n+1-i})\;\L^{(i)}\,,$
and $Q\,$ is the $s\ell_n\,$ root lattice; the lowest weight
(LW) subspaces are one-dimensional:
\be
\lb{dim1}
{\rm dim} \,{\cal V}_p^{(\L_-)}\;
= 1
\,.
\ee
We shall keep (\ref{dim1}) as a defining property of
${\cal V}_p\,$ even when
${\cal V}^{(\L_-)}_p\,$
is not a LW subspace, i.e. for
$F_i\, {\cal V}^{(\L_-)}_p\,\not\equiv 0\,,$
but the corresponding vector that generates
${\cal V}^{(\L_-)}_p\,$
is cosingular.

\vspace{3mm}

\noindent
ii)~The $U_q(s\ell_n)\,$ Casimir operators are multiples of the identity
in ${\cal V}_p\,;$ their eigenvalues are expressed as polynomials
in $\bq^{p_i}\,$ (coinciding with the Casimir eigenvalues for
finite dimensional irreducible subrepresentations). In particular,
the second order Casimir operator and its eigenvalue are related to
the $U_q(s\ell_n)\,$ invariant $q$-trace of the corresponding
monodromy matrix (\ref{mon}), ${\rm tr}(\bq^{2{\rho^\vee}} M)\,$
(cf. \cite {FRT}), where $\bq^{2\rho^\vee}
:= \bq^{\sum_{\a >0} H_\a}\,$ is taken in the fundamental
representation, $\bq^{2\rho^\vee}
= {\rm diag}\, \{q^{1-n}, q^{3-n},\dots ,q^{n-1}\,\}$:
\be
\lb{Casn}
\{\,{\rm tr}\, (\bq^{2\rho^\vee}\, M) \, -\, \sum_{i=1}^n \bq^{2p_i}\,\}
\,{\cal V}_p \, =\, 0.
\ee
For $n=2\,$ we obtain $( p \equiv p_1-p_2)\,$
\be
\lb{Casn2}
\{\, (\bq - q)^2 FE +\bq^{H +1} + q^{H +1} - q^p - \bq^p\}\,
{\cal V}_p = 0\,,
\ee
an equality that can be also cast into the more familiar form
\be
\lb{Casn2a}
\left( C_2 \, -\, 2 \left[ p-\frac{1}{2} \right]
\left[ p+\frac{1}{2} \right]   \right)
\, {\cal V}_p\, =\, 0\,,\qquad C_2
\, =\,  EF + FE + [2] \left[ {H\over 2} \right]^2 \,.
\ee
For $n=3\,$ (\ref{Casn}) yields the following analogue of (\ref{Casn2}):
\ba
\lb{Casn3}
&&\{ (\bq -q)^2 \left(
F_1\, E_1\bq^{\frac{H_1 +2H_2}{3}+1}+
F_{12}\, E_{21}q^{\frac{H_2 -H_1}{3} -1}+
F_2\, E_2q^{\frac{2H_1 +H_2}{3}+1} \right) +\nonumber\\
&&\\
&&+\,\bq^{\frac{2}{3} (2H_1+H_2 ) +2}+
q^{\frac{2}{3} (H_1 - H_2 )}+
q^{\frac{2}{3} (H_1 + 2H_2 ) +2}
- \bq^{2p_1}-\bq^{2p_2}-\bq^{2p_3}\}\, {\cal V}_p = 0\,.\nonumber
\ea

\vspace{5mm}

The {\em dual space} ${\cal V}'_p\,$ -- i.e. the space of linear
forms ${\cal h} f\,,\,.\,{\cal i}\,$ on ${\cal V}_p\,,$ carries the
contragradient representation $X\,\rightarrow \,\check X\,.$
We shall define it by
\be
\lb{dual}
{\cal h} {\check X} f\,,\, v {\cal i}\, =\,
{\cal h} f\,,\,\gamma\circ\sigma\, (X) v {\cal i}\ \qquad
(f\in {\cal V}'_p\,,\ v\in {\cal V}_p\; )\,,
\ee
where $\g\,$ is the antipode (\ref{gamma}) and $\sigma\,$ a $U_q\,$
automorphism (introduced explicitly for reasons that will be made
clear below).

We shall make use of the following result (cf. Section 3 of
\cite{FHT3}).
\vspace{5mm}

\noindent
{\bf Proposition 2.1}~{\em The associative algebra $U_q(s\ell_n)\,$
admits, for $q\,$ on the unit circle, a linear antiinvolution
$X\rightarrow X'\,$ and an antilinear hermitean conjugation
$X\rightarrow X^*\,$ (both preserving the commutation relations)
which make the Cartan generators $q^{H_i}\,$ symmetric, resp. unitary,
\be
\lb{Cartan}
\left( q^{\pm H_i}\right)' = q^{\pm H_i}\,,\qquad
\left( q^{\pm H_i}\right)^* = q^{\mp H_i}\quad
\ee
and extend to a coalgebra homomorphism, resp. antihomomorphism,
\be
\lb{conj}
\left( X\otimes Y\right)' = X'\otimes Y'\,,\quad
\left( X\otimes Y\right)^* = Y^*\otimes X^*\,.
\ee
The hermitean conjugation is determined from these properties
uniquely:
\be
\lb{*}
E_i^* = F_i\,,\qquad F_i^*=E_i\,.
\ee
The ``transposition" $X\rightarrow X'\,$ is determined by
(\ref{conj}) and (\ref{Cartan}) up to a cyclic inner automorphism,
$E_i \rightarrow q^m E_i\,,\ F_i \rightarrow \bq^m F_i\,,\ m\in
\Z\,$. We shall fix this freedom setting, for the Chevalley
generators,
\be
\lb{'}
E_i' = F_i\, q^{H_i-1}\,,\qquad F_i' = \bq^{H_i-1} E_i\,,
\ee
a choice yielding a symmetric monodromy matrix.}
\vspace{5mm}

We shall choose in what follows the automorphism $\sigma\,$
so that
\be
\lb{sigma}
\g\circ\sigma (X) \, = \, X' \,,\quad\forall X\in U_q\,.
\ee
Using (\ref{gamma}) and defining the transposition as in
(\ref{'}), one can see that for $U_q(s\ell_n)\,$ this
amounts to set $\sigma\,$ equal to the involutive
automorphism
\be
\lb{sigma_n}
\sigma (q^{H_i}) = \bq^{H_i}\,\quad \sigma (E_i) = -q F_i\,,\quad
\sigma (F_i) = -\bq E_i\quad\Rightarrow\quad \sigma^2 = {\rm id}\,.
\ee
The most important feature of the choice (\ref{sigma}) is that it
makes the map $\g\circ\sigma\,$ involutive too, $(\;\g\circ\sigma\;
)^2 = {\rm id}\,;$ the invariance of the pairing on ${\cal V}'
_{p} \times {\cal V}_p\,$ can be now expressed as
\be
\lb{invpair}
{\cal h} \,{\check X} f\,,\, v \,{\cal i}\,=\, {\cal h} \, f\,,\,
X' v\,{\cal i}\quad {\rm for\ \, all}\quad f\in {\cal V}'_{p}\,,\
v\in {\cal V}_p\,,\ X\in U_q\,.
\ee

As it will become explicit in the following sections (see also
\cite{BMP}), in fact the
contragradient representation of ${\cal V}_p\,$ is
{\em equivalent} to that of weight
\be
\lb{p'def}
p'= w_L p =\{ p_n,\dots ,p_1\}\quad
\Rightarrow \quad p'_{i\; i+1}=- p_{n-i\; n-i+1}\,,
\ee
$w_L\,$ being the longest Weyl group element,
\be
\lb{wL}
w_L = w_1 \dots w_{n-1}\, w_1 \dots w_{n-2}\,\dots\, w_1 w_2\, w_1\,,
\ee
i.e. ${\cal V}'_p\,\simeq{\cal V}_{p'}\,$. We note that the involutive
element $w_L\ \, (w_L^2 = 1 )\,$ only coincides with the reflection
$w_\theta\,$ with respect to the highest root $\theta\,$ for $n=2,3\,.$
Identifying $w_L\,$ with its $(n-1)$-dimensional representation in the
basis of fundamental weights $\Lambda^{(i)}\,$ we find that ${\rm det}\,
w_L = (-1)^{n\choose{2}}\,.$

The counterpart of the weight space expansion (\ref{wVp}) for
${\cal V}_{p'}\,$ reads
\ba
\lb{wV'p}
{\cal V}_{p'} = {\oplus}_{\lambda \in {\L_-}' + Q} {\cal
V}_{p'}^{(\lambda )}\,,\ \quad {\rm dim}\,{\cal V}_{p'}^{(\lambda )}
\, =\,{\rm dim}\,{\cal V}_p^{(\lambda )}\,,
\ea
the (finite dimensional) space ${\cal V}_{p'}^{(\lambda )}\,$ being
equivalent to the dual to ${\cal V}_p^{(\lambda )}\,.$ It follows from
(\ref{dual})
that for homogeneous elements $v_{\lambda}\in {\cal V}_p^{(\lambda
)}\,$ and $f_\mu\in {\cal V}_{p'}^{(\mu )}\,$ the pairing ${\cal h}
f\,,\, v {\cal i}\,$ is only nonzero if $\mu =
\lambda\,:$
\be
\lb{formnon0}
{\cal h} f_\mu\,,\, v_\lambda {\cal i} = \d_{\mu \lambda }
\,{\cal h} f_{\lambda}\,,\, v_{\lambda} {\cal i}\,.
\ee

\vspace{5mm}

\noindent
{\bf Remark 2.2}~
Using the notation ${\cal V}_{p'}\,$ we thus extend the admissible values
of $p\,$ to Weyl group images of dominant weights which are no longer
dominant. We shall adopt the resulting more general label for the class of
indecomposable representations of interest in what follows.

\vspace{5mm}

For $p_{1n} < h\,$ there is a unique finite dimensional {\em
irreducible} $U_q(s\ell_n)\,$ module ${\cal V}_p\,$ and its dual is
the (irreducible) contragradient module ${\cal V}_{p'}\,.$
If we allow for indecomposable modules, there is of course a wider
list of possibilites
even when one restricts attention, as we shall, to finite
dimensional representations. In this more general case
dual pairs $\left({\cal V}_p\,,\,{\cal V}_{p'}\,\right)\,$ are
characterized by the following relationship among their singular
and cosingular vectors.
\vspace{5mm}

\noindent
{\bf Proposition 2.3}~{\em There is a duality between singular
vectors and equivalence classes of cosingular vectors:
\be
\lb{singcosing}
{\rm Ker}\, {\left( U_q^{\mp} |_{{\cal V}_p}\right)}' \, =\,
{\cal V}_{p'} / U_q^{\pm}{\cal V}_{p'}\,.
\ee
In other words, the bilinear pairing ${\cal h}~ , ~{\cal i} :\,
{\cal V}_{p'}\times {\cal V}_p \,\rightarrow\, \C\,$ projects to
nondegenerate pairings
\be
\lb{ndg}
 {\cal h}~ , ~{\cal i} :\,
{\cal V}_{p'} / U_q^{\pm}{\cal V}_{p'}\,\times
{\rm Ker}\, U_q^{\mp} |_{{\cal V}_p} \,\rightarrow \, \C\,.
\ee
}
\vspace{2mm}

\noindent
{\bf Proof}~The argument is the same as in Section 4 of \cite{BF}.
To fix the ideas, we shall demonstrate that $ {\rm Ker}\, U_q^-
|_{{\cal V}_p}\,$ is the subspace of ${\cal V}_p\,$ orthogonal to
$U_q^+ {\cal V}_{p'}\,.$ Let $v\in
\left( U_q^+ {\cal V}_{p'}\right)^\perp\, (\,\subset {\cal V}_p\,
)\,;$ then, for all $f\in {\cal V}_{p'}\,,\ X^+ \in U_q^+\,$
\be
\lb{perp}
0\, =\, {\cal h} {\check X}^+ f\,,\, v{\cal i}\, = \, {\cal h} f\,,\,
\g\circ\sigma (X^+ ) v{\cal i}\ \Rightarrow\ U_q^- v\, =\,0\,.
\ee
(In the last implication we have used the nondegeneracy of the form
${\cal h}~ , ~{\cal i} \,$ and the fact that the map $
\g\circ\sigma\,
: \, U_q^+\,\rightarrow\, U_q^-\,$ is onto\,.) This proves that
each vector $v\in \left( U_q^+ {\cal V}_{p'}\right)^\perp\, \subset
{\cal V}_p\,$ is singular. The converse statement follows from the
fact that the map $\g\circ\sigma\,$is invertible.
\vspace{5mm}

\section{Indecomposable $h$-dimensional representations of
$U_q(s\ell_2)\,$ }

\medskip
\setcounter{equation}{0}
\renewcommand{\theequation}{\thesection.\arabic{equation}}

We shall study in this Section the simplest,
rank $r=1,\,$ case of $U_q(s\ell_2)\,$ generated by $E, F, q^H,
\bq^H$ satisfying $q^H \bq^H = 1 = \bq^H q^H\,$ and
\be
\lb{n=2}
[E\,,F] = [H] = {{q^H - \bq^H}\over{q-\bq}}\,,\quad
q^H E = E q^{H+2}\,,\quad
q^H F = F q^{H-2}\,.
\ee
A major simplification in this case results from the fact that both
products $E F\,$ and $F E\,$ are expressed in terms of the Cartan
generators $q^{\pm H}\,$ and the
central operator $\left[{p\over 2}\right]^2\,$ defined in terms of
the quadratic Casimir invariant (\ref{Casn2a}), or
equivalently,
\be
\lb{Casn2c}
EF + \left[{{H-1}\over 2}\right]^2 =
\left[{p\over 2}\right]^2 = FE + \left[{{H+1}\over 2}\right]^2\,.
\ee

There are three inequivalent dual pairs of $h$-dimensional
representations of $U_q(s\ell_2)\,$ corresponding to different
``rational factorizations" of $EF\,$ and $FE\,$. All six
representations share the existence of a canonical
weight basis $v_{p m}\,$ such that
\be
\lb{v1}
(q^H - q^{2m-p+1}) v_{p m} = 0
\ee
\be
\lb{v2}
(EF - [m][p-m]) v_{p m} = 0 = (FE - [m+1][p-m-1])v_{p m}\,.
\ee
Clearly, the eigenvalues of $q^H, EF\,$ and $FE\,$ are periodic
in $m\,$ (with period $h\,$) and in $p\,$ (with period $2h\,$).
We shall require that the action of $E\,$ and $F\,$ in the six
representations we are going to define preserves this periodicity
property.

First, we introduce the lowest (resp. highest) weight Verma modules
${\cal V}_p^-\ \ (\,{\cal V}_p^+\,)$ for which the action of $E\,$
and $F\,$ on the respective canonical bases $v_{p m} = | p,
m{\cal i}_-\ (\, |p, m{\cal i}_+\,)\,$ are given by
\be
\lb{V-}
E|p, m{\cal i}_-=|p, m+1{\cal i}_-
\ \Rightarrow\
F|p, m{\cal i}_-=[m][p-m]|p, m-1{\cal i}_-\,,
\ee
\be
\lb{V+}
F|p, m{\cal i}_+= |p, m-1{\cal i}_+
\Rightarrow\
E|p, m{\cal i}_+ =[m+1][p-m-1]|p, m+1{\cal i}_+\,.
\ee
We assume that the lowest weight vector in ${\cal V}^-_p\,$ is $|p,
0{\cal i}_-\,,$ and the highest weight vector in ${\cal V}^+_p\,$
is $|p, h-1{\cal i}_+\,.$
Note that the coefficients for both ${\cal
V}^-_p\,$ and ${\cal V}^+_p\,$ are invariant with respect to the
transformation $(p, m)\,\rightarrow\, (-p, m-p)\,$. It is easy to
prove that the representation ${\cal V}_p^+\,$ is equivalent to the
dual of ${\cal V}_p^-\,:$ in the basis $f_{pm}\,$ of $({\cal V}_p^-
)'\,$ defined by ${\cal h} f_{pm} , v_{pn} {\cal i} =
\d_{mn}\,$ one obtains from (\ref{invpair})
$$\check E f_{pm} = q^{2-p+2m} [m+1][p-m-1] f_{p\;m+1}\,,\quad
\check F f_{pm} = q^{p-2m} f_{p\;m-1}$$
(cf. (\ref{V+})), and it only remains to renormalize the basis
vectors $f_{pm}\,\rightarrow\,q^{m(m+1-p)} f_{pm}\,$ to obtain a
full coincidence with (\ref{V+}).

Using the periodicity property in $m\,,$ we can introduce the {\em
cyclic} $h$-dimensional counterparts ${\cal C}^{\pm}_p\,$ of
${\cal V}^\pm_p\,$
identifying all $|p, m{\cal i}_\pm\,$ with $m \ {\rm mod}\ h\,.$

For the second pair ${\cal C}_{\pm p}\,$ of cyclic representations
the canonical basis vectors
$(\, v_{p m} \,=\, )\ |p, m{\cal i}\in {\cal C}_p\,$ and
$|-p, m-p{\cal i}\in {\cal C}_{-p}\,$ both satisfy ((\ref{v1}),
(\ref{v2}) and)
\be
\lb{Vp}
E |p, m{\cal i} =(m+1)_+ |p, m+1{\cal i}\,,\ \
F|p, m{\cal i}=q^{2-p}(p-m)_+|p, m-1{\cal i}\,,\\
\ee
and the counterpart of these relations for negative $p\,$
(reflecting the symmetry $p\leftrightarrow -p\,$
in (\ref{Casn2a}), (\ref{Casn2c}))
\ba
\lb{V-p}
&&E \, |-p, m-p{\cal i}=(m+1-p)_+ |-p, m+1-p{\cal i}\,,\nonumber\\
&&F\, |-p, m-p{\cal i}= - \, q^p\, (m)_-|-p, m-1-p{\cal i}\,.
\ea
Note that one can obtain the $U_q(s\ell_2)\,$ action in ${\cal C}_p\,$
from that in ${\cal C}_p^-\,$ by a simple renormalization of the
canonical basis vectors defining $|p, m{\cal i} := {1\over{(m)_+!}}
|p, m{\cal i}_-\,$ for $0\le m\le h-1\,$. The corresponding
extension of this transformation to the Verma module would be
singular.

The fact that ${\cal C}_{-p}\,$ is equivalent to the dual of ${\cal
C}_p\,$ follows again from an explicit calculation. Indeed, for $v_{pm}
= |p, m{\cal i} \in {\cal C}_p\,$ and $f_{pm} \in ({\cal C}_p )'\,$ such
that ${\cal h} f_{pm} , v_{pn} {\cal i} =
\d_{mn}\,,$ applying (\ref{invpair}) one finds
$$ \check E f_{pm} = - q^2 (m+1-p)_+ f_{p\;m+1}\,,\quad
\check F f_{pm} = q^{p-2} (m)_- f_{p\;m-1}\,.$$
To see the equivalence of this law with (\ref{V-p}), it is enough to
renormalize $f_{pm} = (-q^2)^m | -p , m-p{\cal i}\,.$

We shall be chiefly interested in the third pair of modules whose
existence exploits the fact that for vanishing $EF\,$ and $FE\,$ there
is an additional freedom in choosing $E\,$ or $F\,.$ We define from the
outset the module ${\cal V}_p\,$ as an $h$-dimensional vector space
spanned by $\{ |p, m {\cal i}\,,\ 0\le m\le h-1\,\}\,$ such that
$|p, 0{\cal i}\,$ is a LW vector and $|p, h-1{\cal i}\,$ is a HW
vector. The basis vectors are assumed to satisfy (\ref{Vp}) except that
$F |p, 0{\cal i} = 0\,.$
A natural basis for displaying its properties is given by
$$
e_{p m} = ([m]!)^{-\frac{1}{2}} E^m | p, 0{\cal i}\quad (\, =
([m]!)^{\frac{1}{2}} q^{{m\choose 2}} |p, m{\cal i}\,)
$$
yielding
\be
\lb{natural}
E\, e_{p m} = [m+1]^{\frac{1}{2}} e_{p\, m+1}\,,\quad F e_{p m} =
[m]^{\frac{1}{2}} [p-m] e_{p\, m-1}\,.
\ee
The change of basis $|p, m{\cal i}\ \rightarrow\ e_{p m}\,$ is
non-singular in ${\cal V}_p\,$ since $[m]!\,$ is positive for $0\le m\le
h-1\,.$ However the action of $F\,$ implied by (\ref{natural}) would be only
recovered from (\ref{Vp}) if we replace the periodicity condition $|p,
m+h{\cal i} = |p, m{\cal i}\,$ by the boundary condition $|p, -1{\cal i} =
0\,.$ Alternatively, we can set
\be
\lb{Fp}
F|p, m{\cal i}=(1-\d (m)) q^{2-p}(p-m)_+|p, m-1{\cal i}\,,\ \
E |p, m{\cal i} =(m+1)_+ |p, m+1{\cal i}\,;
\ee
then ${\cal V}_p\,$ can be continued to a periodic module, too,
by extending
the Kronecker $\d\,$ periodically (setting $\d (m) = 1\,$ for $m=0\
{\rm mod}\, h\,,\ \d (m)=0\,$ otherwise). Such an extension would
result, however, in just a direct sum of infinitely many equivalent
$h$-dimensional $U_q(s\ell_2)\,$ submodules.
It is the realization (\ref{Fp}) which justifies preserving the term
``rational factorization" for the module ${\cal V}_p\,.$ It will prove more
convenient to use (\ref{Fp})
in the description of dual modules and ``BRS operators"
below. The existence of a non-singular basis satisfying (\ref{natural})
makes manifest the consistency of using the artificially looking factor
$(1-\d (m))\,$ in the action of $F\,$ (\ref{natural}).
The module ${\cal V}_p\,$ for $0<p<h\,$ is
dual to ${{\cal V}}_{2h-p}\,, (\simeq {{\cal V}}_{-p})\,$
where ${{\cal V}}_{2h-p}\,$ can be defined as follows:
\ba
\lb{calFp}
&&{{\cal V}}_{2h-p}\,=\,\{ |2h-p, h+m-p{\cal i}\,,
\ 0\le m\le h-1\,\}\,,\nonumber\\
&&E |2h-p, h+m-p{\cal i} =\nonumber\\
&&= (1-\d (h-m-1)) (m+1-p)_+ |2h-p, h+m-p+1{\cal i}\,,\nonumber\\
&&F|2h-p, h+m-p{\cal i}= - q^p (m)_-
|2h-p, h+m-p-1{\cal i}\,.
\ea
(The range of values for $m\,$ corresponds to the target space of the BRS
operator $Q^{h-p}\,$ defined below, see (\ref{Qh-p}).)
Although Eqs. (\ref{Fp}) and (\ref{calFp}) make sense for $0<p<2h\,,$ we
shall assume $0<p<h\,$ for both ${\cal V}_p\,$ and ${\cal V}_{2h-p}\,,$
thus avoiding ambiguity. Note that for $p=h\,$ the two dual modules become
equivalent and irreducible.

\vspace{5mm}

\noindent
{\bf Remark 3.1~} The modules ${\cal C}_p\ (p\in \Z\,)\,$ are, unlike
${\cal C}_p^\pm\,$ and $({\cal V}_p\,,\, {{\cal V}}_{2h-p} )\,,$
self-conjugate in the following sense. There exists an
antiunitary operator $\T\,$ in ${\cal C}_p\,$ of unit square
which implements the
hermitean conjugation (\ref{*}). It acts on the canonical basis according
to
\be
\lb{T}
\T\, | p , m {\cal i}\, = \, q^{m(m+1-p)}\, | p , p - 1 - m {\cal i}\,.
\ee
In verifying its properties we use antilinearity of $\T\,;$ e.g.
\ba
&&\T E \T |p , m {\cal i} = \T q^{m(m+1-p)} (p-m)_+ | p , p-m {\cal
i} = \nonumber\\
&&= \bq^{m(m+1-p)} {\overline{(p-m)_+}} q^{(p-m)(1-m)} | p , m-1 {\cal
i}
= q^{2-p} (p-m)_+ | p , m-1 {\cal i} = \nonumber\\
&&= F | p , m {\cal i}\,,\nonumber\\
&&\T q^H \T |p, m {\cal i} = q^{m(p-1-m)} \T q^{p-1-2m}
| p, p-1-m {\cal i}
= q^{2m-p+1} | p , m {\cal i} =\nonumber\\
&& =  q^H | p , m {\cal i}\,.
\lb{TET}
\ea
In particular, $\T\,$ exchanges the HW vector
$|p , h-1 {\cal i}\,$ with the LW one, $|p , p {\cal i} \
(\equiv |p , p-h {\cal i}\, )\,.$
There is no operator with these properties mapping
${\cal C}_p^+\ ( {\cal C}_p^-\,)\,$
into itself since
${\cal C}_p^+\ ( {\cal C}_p^-\,)\,$
only admits highest (lowest) weight vectors. One can, however, define a
$\T\,$ which intertwines ${\cal C}_p^+\,$ with
${\cal C}_p^-\,.$

\vspace{5mm}

The representations described above are characterized by the
structure of their singular and cosingular vectors. The cyclic
module ${\cal C}_p^-\,$ contains two singular vectors, $|p, 0{\cal
i}_-\,$ and $|p, p{\cal i}_-\,$ and no cosingular ones, whereas in
${\cal C}_p^+\,$ there are two cosingular, $|p, 0{\cal i}_+\,$ and
$|p, p{\cal i}_+\,$, and no singular vectors. Both representations
are therefore irreducible.

The $h$-dimensional cyclic representations
${\cal C}_{\pm p}\,$ and the modules
${\cal V}_p\,, \ {{\cal V}}_{2h-p}\,$
are, on the other hand, indecomposable.
Namely, ${\cal C}_p\,$ and ${\cal V}_p\,$ admit an
$(h-p)$-dimensional invariant subspace ${\cal I}_{h-p}\,$ while
${\cal C}_{2h-p}\,$ and ${{\cal V}}_{2h-p}\,$ contain a
$p$-dimensional invariant subspace
${\cal I}_p\,$. The lowest and highest weight vectors of the
(irreducible) invariant
subspaces of ${\cal C}_p\,$ and  ${\cal C}_{2h-p}\,$ are
\be
\lb{LHWV}
\{\, |p,p{\cal i}\,,\ |p,h-1{\cal i}\,\}\, \in {\cal
I}_{h-p}\,,\ \
\{\, |2h-p,h-p{\cal i}\,,\ |2h-p,h-1{\cal i}\,\}\, \in
{\cal I}_p\,,
\ee
respectively; the vectors $| \pm p , \pm p {\cal i}\,$
are the only singular
vectors in ${\cal C}_{\pm p}\;$, and $|\pm p, 0{\cal i}\,$ are the
cosingular ones. The labels of ${\cal C}_{\pm p}\;$ can be given
now the following meaning: they are equal to the eigenvalue (mod
$2h$) of the operator $1-H\,$ on the cosingular vector or,
equivalently, of $H-1\,$ applied to the corresponding singular
vector (cf. Eqs. (\ref{v1}) and (\ref{LHWV})).

It follows from (\ref{Vp}) and (\ref{V-p}) that the operators $E\,$
and $F\,$ are nilpotent in both ${\cal C}_{\pm p}\,,$ and ${\cal
V}_p\,,\ {{\cal V}}_{2h-p}\,;$ e.g.
\be
\lb{nil}
E^h\, =\, 0\, = F^h\ (\, = q^{hH} - (-1)^{p-1}\,)\,,
\ee
and identical relations are true for $\check E\,,\ \check F\,,
\ q^{\check H}\,$.

The irreducible $U_q(s\ell_2)\,$ representation in ${\cal I}_{h-p}\,$
is equivalent to the one in the quotient
${\cal C}_{2h-p}/{\cal I}_p\,$ or
${{\cal V}}_{2h-p}/{\cal I}_p\,$
and vice versa. The partial equivalence of
${\cal C}_p\ $ (${\cal V}_p\,$) with
${\cal C}_{-p}\simeq{\cal C}_{2h-p}\ $
( ${{\cal V}}_{-p}\simeq{\cal V}_{2h-p}\,$)
will be displayed by explicitly constructing the
corresponding intertwining maps $Q^p\,$ and $Q^{h-p}\,$
which can be characterized by their invariance properties
\be
\lb{invQ}
X Q^p = Q^p {\check X}\,,\quad {\check X} Q^{h-p} = Q^{h-p} X\,,
\ \ \forall X\in U_q(s\ell_2)\,.
\ee

\noindent
{\bf Proposition 3.2}~{\em The maps
$Q^p : {\cal C}_{-p}\rightarrow {\cal C}_p\,$
and $Q^{h-p} : {\cal C}_p\rightarrow {\cal C}_{2h-p}\,$
are determined up to $m$-independent factors from the
$U_q(s\ell_2)\,$ invariance conditions (\ref{invQ}) and are given by
\be
\lb{Qp}
Q^p\, |-p,m-p{\cal i} = {{(m)_+!}\over{(m-p)_+!}}\, |p,m{\cal i}\,,\\
\ee
\be
\lb{Qh-p}
Q^{h-p}\, |p,m{\cal i} = {{(h+m-p)_+!}\over{(m)_+!}}\, |2h-p,h+m-p{\cal
i}\,.
\ee
They satisfy
\be
\lb{cohom2}
{\rm Ker}\, Q^p = {\cal I}_p = {\rm Im} \,Q^{h-p}\quad
(\,{\cal C}_{-p}\simeq{\cal C}_{2h-p}\,)\,,\qquad
{\rm Im}\, Q^p = {\cal I}_{h-p} = {\rm Ker}\, Q^{h-p}\,.
\ee
Furthermore, $Q^p\,$ is the $p$-th and $Q^{h-p}\,$ -- the
$(h-p)$-th power of an operator $Q\,$ acting on
$\oplus_{p\in\Z}{\cal C}_p\,$ as
\be
\lb{basicQ}
Q |p,m{\cal i} = (m+1)_+ |p+2,m+1{\cal i}\,.
\ee
}

The BRS operators (\ref{Qp}) and (\ref{Qh-p}) also act on
${\cal V}_p\,,\ {{\cal V}}_{-p}\,:$
\be
Q^p :\ {{\cal V}}_{-p}\ \to \ {\cal V}_p\,,\quad
Q^{h-p} :\ {\cal V}_{p}\ \to \ {{\cal V}}_{2h-p}\,.
\ee

Alternatively, one can define the intertwiners between the dual pairs
of representations using the following fact.

\vspace{5mm}

\noindent
{\bf Proposition 3.3~}{\em There exist (unique up to normalization)
bilinear forms $(\Phi | \Psi )\,$ on ${\cal C}_{\pm p}\,$ and on
${\cal V}_{\pm p}\,$
for $0<p< h\,$ invariant with respect to the transposition
$X\,\rightarrow\, X'\,$ of Proposition 2.1,}
\be
\lb{invbil}
(\,\Phi\, |\, X\; \Psi\, )\, =\, (\, X'\; \Phi\, |\, \Psi \, )\,.
\ee
{\em The form on ${\cal C}_p\,$ and on ${\cal V}_p\,$
vanishes on the $(h-p)$-dimensional
$U_q(s\ell_2)\,$ invariant subspace ${\cal I}_{h-p}\,$ with lowest
and highest weight vectors $|p,p{\cal i}\,$ and $|p,h-1{\cal i}\,$,
respectively, while
\ba
\lb{Np}
&&(p,m |p, m') = {{N_p \,\d_{mm'}}\over{(m)_+!(p-m-1)_-!}}\nonumber\\
&&{\rm for}\ 0\le m\le p-1\ \ {\rm or}\ \ 0\le m'\le p-1.
\ea
It satisfies the conjugation property
\be
\lb{conj1}
\overline{(p,m |p, m)} = (p,p-m-1 |p, p-m-1)\ {\rm for}\
N_p={\overline{N_p}}\,.
\ee
The corresponding relations for ${\cal C}_{-p}\,$ and
${{\cal V}}_{-p}\,$ are
\ba
\lb{N-p}
&&(-p, m-p |-p, m'-p) = {{N_{-p}\,
\d_{mm'}}\over{(h-1-m)_-!(m-p)_+!}}\nonumber\\
&&(\, = 0 \ {\rm for}\ 0\le m\le p-1 \ \ {\rm or}\ \ 0\le m'\le
p-1\,, \ {\rm i.e. on}\  \ {\cal I}_p )\,,
\ea
and, for $p\le m\le h-1\,$,
\ba
\lb{conj2}
&&\overline{(-p, h-1-m |-p, h-1-m)} = \nonumber\\
= &&(-p,m-p |-p, m-p)\ {\rm for}\  N_{-p}={\overline{N_{-p}}}\,.
\ea
}
(Note that in (\ref{invbil}), in contrast to (\ref{invpair}),
$X'\,$ is in the {\em {same}} representation as $X\,,$
not in the contragradient one.)

\vspace{3mm}

\noindent
{\bf Proof~} The orthogonality for $m\ne m'\,$ follows from
$q^H\,=\, (q^H )'\, $. Applying (\ref{invbil}) say, for $\Phi = |p,
m+1 {\cal i}\,, \ \Psi = |p, m {\cal i}\,$ (resp., $\Phi = | -p,
m-p+1 {\cal i}\,, \ \Psi = |-p, m-p {\cal i}\,)$ and $X= E\,,$ one
obtains recurrence relations solved by (\ref{Np}) and (\ref{N-p}),
respectively.
\vspace{5mm}

The (degenerate) invariant bilinear forms on ${\cal C}_p\,$ and
${\cal C}_{-p}\,$ are related, through the intertwiners $Q^{h-p}\,$ and
$Q^p\,$, to the pairing between dual modules.
One can prove the following relations that can be used
alternatively as a definition of the intertwiners:
\be
\lb{Q}
{\cal h} f_1 \,,\, Q^p f_2 {\cal i} \, =
\, M_{-p}\, ( \, f_1\, |\, f_2\, )\,,\quad
{\cal h} Q^{h-p} v_1 \,,\, v_2 {\cal i}\, =
\, M_p\, ( \, v_1\, | \, v_2\, )\,,
\ee
where $v_1, v_2\in {\cal C}_p\,,\ f_1, f_2 \in {\cal C}_{-p}\,$ and
$M_{\pm p}\,$ are $m$-independent.

We can now prove the equivalence of (\ref{invQ})
and (\ref{Q}):
\ba
\lb{alt}
&&{\cal h} \, f_1 ,\, X \, Q^p\, f_2\, {\cal i} =
{\cal h} \,{\check X}'\, f_1\,,\, Q^p\, f_2\, {\cal i} =\\
&&= M_{-p} \,(\, {\check X}'\, f_1\, |\, f_2\, ) =
M_{-p}\, (\, f_1\, |\, {\check X}\, f_2\, ) =
{\cal h}\, f_1\,,\, Q^p\, {\check X}\, f_2\, {\cal i}\,,\nonumber\\
&&{\cal h}\,{\check X}\, Q^{h-p}\, v_1\,,\, v_2\, {\cal i} =
{\cal h}\, Q^{h-p}\, v_1\,,\, X'\, v_2\, {\cal i} =\\
&&= M_p \,(\, v_1 \, |\, X'\, v_2\, ) =
M_p\, (\, X \, v_1\, |\, v_2\, )
= {\cal h}\, Q^{h-p}\, X\, v_1\,,\,  v_2\,{\cal i}\,.\nonumber
\ea
We have used (\ref{invpair}), (\ref{Q}) and the involutivity of
the transposition (\ref{conj})
(note also that $({X'}){\check{\phantom 1}} = ({\check X})' \equiv
{\check X}'\,)\,.$

From (\ref{Qp}), (\ref{Qh-p}), (\ref{Q}) and (\ref{M}) it follows that
\be
\lb{NN}
{\cal h} -p, m-p \, |\, p, m {\cal i} = (-1)^m q^{1-p+2m}
N_{-p}\equiv (-1)^m (\bq^{2\rho^\vee})^m_m N_{-p}\,,
\ee
\be
\lb{NN1}
{\cal h} 2h
-p, h+m-p \, |\, p, m {\cal i} = (-1)^{m-p} q^{1-p+2m} N_p\equiv
(-1)^{m-p} (\bq^{2\rho^\vee})^m_m N_p
\ee
for
\be
\lb{M}
M_p \, = \,  q^{p-h-1}\,  (h-1)_+!\,,\quad
M_{-p}\, = \,  q^{1-p}\,  (h-1)_-!\,.
\ee
To derive (\ref{NN}) - (\ref{M}) one uses relations (\ref{Qp}),
(\ref{Qh-p}) and
\be
\lb{h-k}
(h-k)_\pm ! =
{{(h-1)_\pm ! (-q^{\pm 2})^{k-1}}\over {(k-1)_\mp !}}\,.
\ee
Note that in the limit $q\,\to \, 1\,$ one gets the well known factor
$(-1)^m\,$ relating the covariant and the contravariant canonical
bases of $s\ell_2\,$ modules (\cite{Ham} 11-6).
Equating the two expressions one obtains the relation
\be
\lb{Nrel}
N_{-p} = (-1)^p N_p\,,
\ee
consistent with the reality of $N_{\pm p}\,$.

\vspace{5mm}

\noindent
{\bf Remark 3.4}~The modules ${\cal C}_p\,$ and ${\cal V}_p\,$ also admit
an invariant hermitean form obtained by using the hermitean
conjugation $X\ \rightarrow\ X^*\,$ of Proposition 2.1 instead of the
transposition. It can be demonstrated to be positive
semidefinite and to majorize the bilinear form. There is a natural choice
of normalization for which the (hermitean) norm square of the basis
vectors is given by
\be
\lb{hermnorm}
\| | p, m{\cal i} \|^2 = | (p,m | p, m )| =
\frac{|N_p|}{[m]![p-m-1]!}\quad{\rm for}\ \ 0<p\le h\,.
\ee

\vspace{5mm}

The following statement allows to interpret the operator $Q\,$ as
a generalized exterior derivative \cite{DV} or ``BRS operator"
\cite{DT} with a trivial cohomology.

\vspace{5mm}

\noindent
{\bf Proposition 3.5}~{\em The operator $Q\,$ (\ref{basicQ}) satisfies
\be
\lb{Qh}
Q^h = 0 = Q^p Q^{h-p} = Q^{h-p} Q^p\,,
\ee
so that the sequence of $U_q(s\ell_2)\,$ modules and intertwiners
\be
\lb{sequence}
\dots
\stackrel{Q^p}{\longrightarrow} \, {\cal C}_{p-2h}\,
\stackrel{Q^{h-p}}{\longrightarrow} \, {\cal C}_{-p}\,
\stackrel{Q^p}{\longrightarrow} \, {\cal C}_p\,
\stackrel{Q^{h-p}}{\longrightarrow} \, {\cal C}_{2h-p}\,
\stackrel{Q^p}{\longrightarrow} \, \dots
\ee
is a complex. Furthermore, it is exact, i.e. all cohomology
groups are trivial (see Eq.(\ref{cohom2})).}
\vspace{5mm}

In fact, the triviality of the cohomologies is, according to
\cite{DV}, a consequence of the existence of an operator $K :
{\cal C}_{p+2} \rightarrow {\cal C}_p\,$ which, together with
$Q\,$ (\ref{basicQ}) satisfies
\be
\lb{K}
K Q - q^2 Q K = \id\,,\quad K^h = 0\,.
\ee
It is given by
\be
\lb{defK}
K |p+2, m+1{\cal i} = (1 - \d_{{h-1\, m}\, ({\rm mod}\, h)})
|p,m{\cal i}\, \quad \left(\Rightarrow\ K|p,0{\cal i} = 0\,\right)\,.
\ee

\vspace{5mm}

\section{$2h$-dimensional indecomposable represen-\\tations of
$U_q(s\ell_2))$}

\setcounter{equation}{0}
\renewcommand\theequation{\thesection.\arabic{equation}}

\medskip

The question we address in this Section is:
can we use the intertwining
map $Q^{h-p}:\ {\cal C}_p\ \to\ {\cal C}_{2h-p}\,$ to
derive raising and lowering
operators ${\frak e}\,$ and ${\frak f}\,$ on
${\cal C}_p\oplus {\cal C}_{2h-p}\,$
of the form
\be
\lb{4.1}
{\frak e} = E+\a (q^{\check H} , q^p) {\check E} Q^{h-p} \oplus {\check
E}\,,\quad
{\frak f} = F + \b (q^{\check H} , q^p ) {\check F} Q^{h-p} \oplus {\check
F}\,,
\ee
$q^{\pm{\frak h}} = q^{\pm (H\oplus\check H )}\,,$
where, for $X\,$ standing for an
endomorphism of ${\cal C}_p\,$ which represents
an element of $U_q(s\ell_2)\,,\ {\check X}\,$
stands for the corresponding
endomorphism of ${\cal C}_{2h-p} ?$ This would equip the direct sum
${\cal C}_p\oplus {\cal C}_{2h-p}\,$ with the structure of an
indecomposable
$U_q(s\ell_2)\,$ module.

It turns out that the problem, as stated, has no solution. Its study,
however, did lead us to a (slightly more general) construction of a
$2h$-dimensional indecomposable $U_q(s\ell_2)\,$
module ${\cal D}_p\,$ such
that
${\cal C}_{2h-p}\,$ appears as an invariant subspace of ${\cal D}_p\,$
while ${\cal C}_p\,$ is isomorphic to the quotient space ${\cal D}_p /
{\cal C}_{2h-p}\,.$ Trying to interpret this result as a realization of the
construction (\ref{4.1}), we would find out that the coefficients $\a\,$
and $\b\,$ are singular functions of their arguments whose poles, however,
are compensated by zeros in the kernel of $Q^{h-p}\,.$ We thus end up with
operators ${\frak e}\,$ and ${\frak f}\,$ of a form suggested but not
literally given
by (\ref{4.1}). Let us reproduce this heuristic argument.

We look for functions $\a_m (p)\,$ and $\b_m (p)\,$ such that the
operators ${\frak e}\,$ and ${\frak f}\,$, acting on the canonical basis
in ${\cal C}_p\,$ as
\ba
\lb{4.2}
&&{\frak e} |p , m{\cal i}= (m+1)_+ |p , m+1{\cal i} + \nonumber\\
&& +\, \a_m (p)\frac{(h+m-p+1)_+!}{(m)_+!}
|2h-p , m-p+1{\cal i}\,,\nonumber\\
&&{\frak f} |p , m{\cal i} = - q^{p+2-2m} ( \, (m-p)_+ |p , m-1{\cal i} +
\nonumber\\
&&+ \b_m (p)\frac{(h +m -p)_+!}{(m-1)_+!} |2h-p , m-p-1{\cal i}
\, )\,,
\ea
satisfy the commutation relation $\left( [{\frak e} , {\frak f} ] -
[ {\frak h} ]
\right)\, {\cal C}_p =
0\,.$ This yields recurrence relations for $\a_m \ (= \a_m (p) )\,$ and
$\b_m\,:$
\be
\lb{4.3}
q^2 (\a_{m-1} + \b_m ) (m)_+ (m-p)_+ =
(\a_m + \b_{m+1} ) (m+1)_+ (m-p+1)_+\,.
\ee
These equations have a singular solution:
\be
\lb{4.4}
\a_m = \frac{\a (p)}{[m+1][m-p+1]}\ \left( = \frac{q^{2m-p} \a
(p)}{(m+1)_+ (m-p+1)_+} \right)\,,\  \b_m = \frac{\b (p)}{[m][m-p]}\,.
\ee
The product of $\a_m\,$ and $\b_m\,$ with the ratio of factorials in
(\ref{4.2}) can, however, be given an unambiguous meaning; we obtain
\ba
\lb{4.5}
&&{\frak e} |p , m{\cal i}= (m+1)_+ |p , m+1{\cal i} + \nonumber\\
&&+ \a (p) q^{2m-p}
\frac{(h+m-p)_+!}{(m+1)_+!} |2h-p , m-p+1{\cal i}\,,\nonumber\\
&&{\frak f} |p , m{\cal i}= q^{2-p}  (p-m)_+ |p , m-1{\cal i} -
\nonumber\\
&&- \b (p)\frac{(h+ m -p-1)_+!}{(m)_+!} |2h-p , m-p-1{\cal i}\,.
\ea
The ratio in the second terms can be defined for a vanishing
denominator
using the following general formula for $q$-binomial coefficients at
roots of $1\,$ (\cite{L}):
\be
\lb{4.6}
{{m_0 +h m_1}\choose{n_0 + hn_1}}_+
={{m_0}\choose{n_0}}_+
{{m_1}\choose{n_1}}_{q=1}
{\rm for}\  m_1\in {\Z} ,\  n_1\in {\Z}_+ ,\
0\le m_0 ,n_0 \le h-1 ;
\ee
this gives, e.g. $\frac{(2h-1-p)_+!}{(h)_+!} \, =\, (h-p-1)_+!\,$
for $0<p <h\,.$
Then the second term in the expression for ${\frak e} |p , m{\cal i}\,$
vanishes for $p\le m \le h-2\,.$
For $p=h\,$ the second term in the right hand side
of both equations (\ref{4.5}) should disappear.
A convenient choice for
$\a (p)\,$ and $\b (p)\,$ which satisfies this condition is
\be
\lb{4.7}
\a (p) = \frac{1}{(h-p-1)_+!} = \b (p)\,.
\ee
Inserting this into (\ref{4.5}) we find
\ba
\lb{4.8}
&&{\frak e}\, |p , m{\cal i}= (m+1)_+\, |p , m+1{\cal i} + \nonumber\\
&&+ q^{2m-p}
{{h+m-p}\choose{m+1}}_+ |2h-p , m-p+1{\cal i}\,,\nonumber\\
&&{\frak f}\, |p , m{\cal i}= q^{2-p}\, (p-m)_+\, |p , m-1{\cal i} -
\nonumber\\
&&- {{h+m-p-1}\choose{m}}_+ |2h-p , m-p-1{\cal i} \,,
\ea
where ${{N}\choose{n}}_+\,$ is the $q$-binomial coefficient
$\frac{(N)_+!}{(n)_+!(N-n)_+!}\,$ which is defined to vanish for $N<n\,.$
Completing (\ref{4.8}) with
\ba
{\frak e}\, |2h-p , m-p{\cal i}&=&
{\check E} |2h-p , m-p{\cal i}=
(m+1-p)_+ |2h-p , m+1-p{\cal i}\,,\nonumber\\
&\phantom{=}
\lb{4.9}
&\\
{\frak f}\, |2h-p , m-p{\cal i}&=&
{\check F} |2h-p , m-p{\cal i}=
- q^p  (m)_- |2h-p , m-p-1{\cal i}\,,\nonumber
\ea
we arrive at the following result.

\vspace{5mm}

\noindent
{\bf Proposition 4.1~} {\em
The $2h$-dimensional vector space ${\cal D}_p\,$ with basis
$$\{ | p, m{\cal i}\in {\cal C}_p \,,\ |2h-p , m-p {\cal i}\in{\cal
C}_{2h-p}\,,
\ m\; {\rm mod} \; h\,
\}\,$$
equipped with the $U_q(s\ell_2)\,$ action 
(\ref{4.8}), (\ref{4.9}) is an indecomposable
$U_q(s\ell_2)\,$ module with the following chain of invariant subspaces:
\be
\lb{4.10}
{\cal I}_p \subset {\cal C}_{2h-p} \subset {\tilde{\cal C}}_{2h-p} \subset
{\cal D}_p\,.
\ee
Here ${\cal I}_p\,$ is the $p$-dimensional invariant
subspace of the $h$-dimensional cyclic submodule ${\cal C}_{2h-p}\,$
with a lowest and a
highest weight vector, $|2h-p , h- p {\cal i}\,$ and
$|2h- p , h-1{\cal i}\,,$ respectively.
${\tilde{\cal C}}_{2h-p}\,$ is spanned by ${\cal C}_{2h-p}\,$ and
${\cal I}_{h-p}\,.$
We further have the following identifications (in
the notation of Section 3)
of the cyclic module ${\cal C}_p\,$ and of the
irreducible subquotient ${\cal C}_p / {\cal I}_{h-p}\,:$
\be
\lb{4.11}
{\cal I}_{h-p}\subset {\cal C}_p \simeq {\cal D}_p / {\cal
C}_{2h-p}\,,\quad {\cal C}_p / {\cal I}_{h-p}
\simeq {\cal D}_p / {\tilde{\cal C}}_{2h-p}\,.
\ee
The same construction applies to the pair $({\cal C}_{-p} , {\cal C}_p
)\,$
which is combined in an indecomposable module ${\cal D}_{-p}\,$
with composition series
${\cal I}_{h-p}\subset
{\cal C}_p\subset {\tilde{\cal C}}_p \subset {\cal D}_{-p}\,$
where ${\tilde{\cal C}}_p\,$ is
$(h+p)$-dimensional and
${\cal C}_{-p} \simeq {\cal D}_{-p} / {\cal C}_p\,.$
}

\vspace{5mm}

\noindent
{\bf Remark 4.2~} There are several inequivalent $2h$-dimensional
indecomposable $U_q(s\ell_2)\,$
modules. For instance, one can choose either
$\a (p)\,$ or $\b (p)\,$ equal to zero. The
modules ${\cal D}_p\,$ are singled out as more symmetric.
Indeed, it is only for ${\cal D}_p\,$ that one can extend the
antiunitary operator $\T\,$ of Remark 3.1 .

\vspace{5mm}

\noindent
{\bf Remark 4.3~}
One can define in a similar way $2h$-dimensional indecomposable
representations ${{\cal W}}_p\,$
combining the pairs
$({\cal V}_p , {{\cal V}}_{2h-p} )\,.$
The Fr\"ohlich-Kerler construction (\cite{FK},
section 5.3) is then reproduced by taking
$\b (p) = 0\,,\ \a (p) \ne 0 \,.$

\vspace{5mm}

The motivation of Ref. \cite{FK} for studying
$2h$-dimensional representations has been their
appearance in the tensor product decomposition
of physical ones. We
are advocating here a different point of view.
Indecomposable $U_q(s\ell_2)\,$ modules
are introduced from the outset
(as counterparts of indecomposable affine Kac-Moody
modules of the chiral current algebra).
The ``physical" representations then appear as
appropriate subquotients. A general study of tensor product
expansions of irreducible representations of $U_q(s\ell_2)\,$
for $q\,$ a root of unity is contained in \cite{Arnaudon}.

\vspace{5mm}

\section{Remarks on the general case. Indecomposable $U_q(s\ell_3)$
modules}

\setcounter{equation}{0}
\renewcommand\theequation{\thesection.\arabic{equation}}

\medskip

One way to write down the canonical basis $\{ \, | p , m {\cal i}\,,\
0\le m\le h-1\,\}\,$ in ${\cal V}_p\,$
that readily generalizes to $U_q(s\ell_n)\,$ consists in acting by
a basis of raising operators of the enveloping
algebra on the LW vector $| p , 0 {\cal i}\,:$
\be
\lb{(m)}
| p , m {\cal i} = E^{(m)}\,
| p , 0 {\cal i}\,,\quad E^{(m)} :=
\frac{E^m}{(m)_+!}\quad (\,{\rm for}\ m < h\,)\,,\quad F | p , 0 {\cal
i} = 0\,.
\ee
For $U_q(s\ell_n)\,$ a straightforward extension of
(\ref{(m)}) is provided by substituting for $\{\,E^{(m)}\,\}\,$ the
Poincar\'e-Birkhoff-Witt (PBW) basis in the subalgebra $U_q^+\,$ of
raising operators. It is labeled by $n\choose 2\,$ quantum numbers
$m_{ij}\,,\ 1\le i\le j\le n-1\,,\,$ the powers of $E_i\, (\, \equiv
E_{ii}\, )\,$ and $E_{ji}\,;$ here $E_{ji}\,$ is defined by continuing
inductively the definition (\ref{Serre}) of $E_{i+1\, i}\,:$
\be
\lb{Eji}
E_{j+1\, i} \,=\, E_{ji}\, E_{j+1}\, -\, q\, E_{j+1}\, E_{ji}\quad
{\rm for}\ 1\le i\le j\le n-2\,.
\ee
The case $n=3\,$
is representative, on one hand, since it shares the main complication
in the passage from $n=2\,$ to $n>2$ -- the appearance of weights of
multiplicity higher than $1\,;$ on the other hand, it still allows an
explicit description since the ``canonical"
\cite{L} (or ``crystal" \cite{K}) basis
consists of monomials in $E_i\,$ just for $n\le 3\,.$ We shall, therefore,
proceed to extending the main results of Section 3 to this
case.

\vspace{5mm}

\subsection{Finite dimensional factor algebra of the Borel subalgebra
$U_q^E\,.$ Lowest weight $U_q(s\ell_3)\,$ modules}

\medskip

$U_q^E\,$ can be viewed as a bigraded associative algebra,
\ba
&&U_q^E = \oplus_{\l_1 ,\l_2 \in \Z_+}^{} U_q^E (\l_1 , \l_2
)  ,\quad U_q^E (\l_1 , \l_2 ) = {\rm
Span\,}\{q^{m_1 H_1 + m_2 H_2} E_1^\a E_{21}^\b E_2^\g\, \}\nonumber\\
&&\a\,,\,\b\,,\,\g\, \in {\Z}_+\,,\
\a+\b = {\l}_1\,,\ \b +\g = {\l}_2\,,\ m_a\in {\Z}\,,
\lb{bigra}
\ea
so that $U_q^E (0,0)\,$ is the group algebra of the Cartan
subgroup of $U_q(s\ell_3 )\,$ generated by $q^{\pm H_a}\,,\ a=1,2\,.$ The linear span of
the {\em PBW basis} $\{E_1^\a E_{21}^\b E_2^\g\, \}\,$ in each $U_q^E (\l_1 , \l_2
)\,$ is taken with operator valued coefficients belonging to $U_q^E (0, 0)\,$ and
satisfying
\be
\lb{l1l2}
q^{H_a}\, U_q^E (\l_1 ,\l_2 )\, \bq^{H_a}\, =\, q^{3\l_a -
\l_1 -\l_2}\, U_q^E (\l_1 ,\l_2 )\,,\ a=1,2\,.
\ee
For each dominant weight ${\bf p} = (p_{12}, p_{23})\,$ we define an
$U_q(s\ell_3 )\,$ module ${\cal V}_{\bf p}\,$ by
\be
\lb{defVbfp}
{\cal V}_{\bf p} = U_h^E |{\bf p}; 0,0,0{\cal i}\,,\ \
U_h^E = \oplus_{\l_1 ,\l_2 = 0}^{h-1} U_q^E (\l_1 ,\l_2 )\,,\ \
F_a |{\bf p}; 0,0,0{\cal i} = 0\,,
\ee
where $|{\bf p};0,0,0{\cal i}\,$ is the LW vector in the PBW basis defined
by
\ba
&&|{\bf p}; \a ,\b ,\g{\cal i} \,
= \, E_1^{(\a )} E_{21}^{[\b ]} E_2^{(\g )}\,
|{\bf p};0,0,0{\cal i}\,,\nonumber\\
&&\a ,\b ,\g \ge 0\,,\quad
\a +\b\le h-1\,,\quad
\b +\g\le h-1\,,
\lb{defPBW}
\ea
where $X^{[\b ]} \equiv \frac{X^\b}{[\b ]!}\,.$ The normalization of the
basis vectors is chosen in such a way that the periodicity of the
eigenvalues of the Cartan generators,
\be
\lb{perCart}
(q^{H_1}-q^{2\a+\b-\gamma -p_{23}+1})
|{\bf p}; \a ,\b ,\g{\cal i}  = 0 =
(q^{H_2}- q^{-\a+\b+2\gamma -p_{12}+1}) |{\bf p}; \a ,\b ,\g{\cal i}\,,
\ee
summed up in the substitution rule
\be
\lb{period}
(\a, \b, \gamma ) \rightarrow (\a+\varepsilon_1 h,
\b+\varepsilon_2 h, \gamma+\varepsilon_3 h)\quad \varepsilon_i = \pm 1\,,\ \
i=1,2,3\,,
\ee
corresponds to the periodicity of the expressions for the action of
$E_a\,$ and $F_a\,\ (a=1,2)\,$ on the PBW basis:
\ba
&&E_1\, |{\bf p}; \a,\b,\gamma {\cal i}\, =\, (\a +1)_+\, |{\bf p};  \a
+1,\b,\gamma{\cal i}\,,\nonumber\\
&&E_2\, |{\bf p}; \a,\b,\gamma {\cal i}\,
=\,\bq^{\a}\, (q^\b\, (\gamma +1)_+\, |{\bf p};  \a,\b,\gamma +1{\cal i} -
\nonumber\\
&&\ - (1-\d_{\a 0})  [\b +1]\, |{\bf p}; \a -1,\b +1,\gamma {\cal i}\,
)\,,\nonumber\\
&&E_{21}\, |{\bf p}; \a,\b,\gamma {\cal i}\, =\, q^{\a}\, [\b +1]\, |{\bf
p};\a,\b+1,\gamma {\cal i}\,,
\lb{E3}\\
&&\nonumber\\
&&F_1\, |{\bf p}; \a,\b,\gamma {\cal i}\, = \bq^{1+\a}
[ p_{23}-\a-\b+\gamma ]\, (1-\d_{\a 0}) |{\bf p}; \a -1,\b,\gamma {\cal
i}+ \nonumber\\
&&\ + q^{\b -p_{23}} [\gamma +1 ] (1-\d_{\b 0}) \, |{\bf p}; \a,\b
-1,\gamma +1 {\cal
i}\,,\nonumber\\
&&F_2\, |{\bf p};
\a,\b,\gamma {\cal i}\, = q^{1-\g} [p_{12}- \gamma ] (1-\d_{\g 0})\, |{\bf p};
\a,\b,\gamma -1{\cal i} -
\nonumber \\
&&\ -\, q^{p_{12}-2\gamma} (\a +1)_+ (1-\d_{\b 0})\, |{\bf p}; \a +1,\b
-1,\gamma
{\cal i}\,,\nonumber\\
&&F_{12}\, |{\bf p}; \a,\b,\gamma {\cal i}\,
=\, \bq^{\a} [p_{13}-\a -\b -\gamma -1 ] (1-\d_{\b 0})\, |{\bf p}; \a,\b -1,\gamma
{\cal i} -
\nonumber\\
&&\ - \,q^{p_{23}+1-2\a -\b} [\gamma -p_{12}]\,
(1-\d_{\a 0}) (1-\d_{\g 0}) |{\bf p};
\a -1,\b ,\gamma -1{\cal i}\,.
\lb{F3}
\ea
The simplest way to verify the consistency of these relations is to use,
following the discussion of the $n=2\,$ case (Eqs. (\ref{natural})), the
non-periodic (but still regular) basis
\be
\lb{ep}
e_{\bf p}(\a ,\b ,\g )
= \left( [\a +\b ]! [\b + \g
]!\right)^{-\frac{1}{2}}\, E_1^\a E_{21}^\b E_2^\g \, |{\bf p}; 0,0,0{\cal
i}\,,
\ee
such that
\ba
&&E_1\, e_{\bf p}(\a ,\b ,\g )
=  \sqrt{[\a +\b ]}\, e_{\bf p}(\a + 1,\b ,\g ) \,,
\lb{E3e}\\
&&E_2 e_{\bf p}(\a ,\b ,\g )
=  \sqrt{[\b + \g ]} ( q^{\b -\a} e_{\bf p}(\a ,\b ,\g +1 )
- \bq [\a ] e_{\bf p}(\a - 1,\b +1,\g ) )\,,\nonumber
\\
&&\nonumber\\
&&F_1\, e_{\bf p}(\a ,\b ,\g )
= \bq^2\,\frac{[\a ]}{\sqrt{[\a +\b ]}}\, [p_{23}-\a -\b +\g ] \,
e_{\bf p}(\a -1 ,\b ,\g )
+ \nonumber\\
&&\ + q^{\b -\g -p_{23}} \frac{[\b ]}{\sqrt{[\a +\b ]}}\, e_{\bf p}(\a ,\b
-1 ,\g +1) \,,
\nonumber\\
&&F_2\, e_{\bf p}(\a ,\b ,\g )
= \frac{[\g ]}{\sqrt{[\b +\g ]}}\, [p_{12}-\g ]\,
e_{\bf p}(\a ,\b ,\g -1) -
\nonumber \\
&&\ -\, q^{p_{12}-\g}\,\frac{[\b ]}{\sqrt{[\b +\g ]}}\, e_{\bf p}(\a +1
,\b
-1 ,\g )\,.
\lb{F3e}
\ea
It is evident that the coefficients in the right-hand sides of
(\ref{F3e})
are defined unambiguously, since the expressions under the square roots
are non-negative, and, for example,
$$ \lim_{\a\to 0} \, \lim_{\b\to 0}\, \frac{[\a ]^2}{[\a +\b ]}
= 0 =  \lim_{\b\to 0} \, \lim_{\a\to 0}\, \frac{[\a ]^2}{[\a +\b ]}\,.
$$

The relations (\ref{E3}), (\ref{F3}), on the other hand, have the
advantage to admit a (periodic) continuation to all integer labels
$\a ,\b ,\g\,.$

In order to find the dimension $d_3(h)\,$ of each ${\cal
V}_{\bf p}\,,$ we compute the dimensions of the corresponding weight
spaces:
\be
\lb{dimweight}
{\cal V}_{\bf p} (\l_1 ,\l_2 ) \,
= \, U_q^E (\l_1 ,\l_2 )\, |{\bf p}; 0,0,0{\cal i}\,,\quad
{\rm dim}\, {\cal V}_{\bf p} (\l_1 ,\l_2 ) \,
=\, {\rm min} (\l_1 , \l_2 ) + 1\,.
\ee
As a result, we find
\be
\lb{d3hPBW}
d_3 (h) := {\rm dim}\,{\cal V}_{\bf p} =
\sum_{\l = 0}^{h-1} (\l +1 ) (2 (h-\l -1 )+1)= \frac{h(h+1)(2h+1)}{6}\,.
\ee

A way to identify the invariant subspaces of ${\cal V}_{\bf p}\,$ passes
through the
construction of singular vectors. As we shall see shortly, the latter need not
belong to the set of PBW basis vectors -- in general, they appear as linear
combinations of basis vectors of a given weight subspace $U_q^E (\l_1 , \l_2 )
|{\bf p}; 0,0,0{\cal i}\,.$
The resulting complication is resolved by passing to the
canonical basis which does contain the LW (and HW) vectors of interest.

The canonical basis in ${\cal V}_{\bf p}\,$ will be defined by acting on
the LW
vector $|{\bf p}; 0,0,0{\cal i}\,$ by two sets of monomials,
$$E_1^{(m)} E_2^{(k)} E_1^{(\ell )}\,,\quad
E_2^{(m)} E_1^{(k)} E_2^{(\ell )}\,,\qquad
m, k, \ell\in{\Z}_+ \,,\quad  k\ge\ell + m\,,$$
which together provide
a basis in $U_q^+\,.$ One
proves (see \cite{L}) that the Serre relations (\ref{Serre}) imply the
following expressions for these monomials in terms of the PBW basis:
\ba
\lb{can-PBW}
q^{k\ell}
E_1^{(m)}E_2^{(k)}E_1^{(\ell )} &=&
\sum_{j=0}^{\ell}
(-1)^j {{m+\ell -j}\choose{m}}_+ E_1^{(m+\ell
-j)}E_{21}^{[j]}E_2^{(k-j)}\nonumber\\
\\
q^{k\ell}
E_2^{(\ell )}E_1^{(k)}E_2^{(m)} &=&
\sum_{j=0}^{\ell}
(-1)^j {{m+\ell -j}\choose{m}}_+ E_1^{(k-j)}E_{21}^{[j]}E_2^{(m+\ell
-j)}\,.\nonumber
\ea
It follows from
(\ref{can-PBW})
that
\be
\lb{121-212}
E_1^{(m)}E_2^{(k)}E_1^{(\ell )} = E_2^{(\ell
)}E_1^{(k)}E_2^{(m)}\quad {\rm for}\ k = m+\ell\,.
\ee
The inverse relations are
\ba
\lb{PBW-can}
E_1^{(\a )}E_{21}^{[\b ]}E_2^{(\gamma )} &=&
\sum_{\sigma =0}^{\b} X_\sigma (\a ,\b , \gamma )
E_2^{(\b -\sigma )}E_1^{(\a +\b )}E_2^{(\gamma +\sigma )}\quad ({\rm for}\
\a\ge\gamma )\,,\nonumber\\
\\
E_1^{(\a )}E_{21}^{[\b ]}E_2^{(\gamma )} &=&
\sum_{\sigma =0}^{\b} X_\sigma (\gamma ,\b ,\a )
E_1^{(\a + \sigma )}E_2^{(\b +\gamma )}E_1^{(\b -\sigma )}\quad ({\rm
for}\ \a\le\gamma )\nonumber
\ea
where
\be
\lb{X}
X_\sigma (\a ,\b ,\gamma ) = (-1)^{\b -\sigma} {{\gamma +\sigma}\choose
\gamma}_+ q^{\sigma (\sigma -1)+(\b -\sigma ) (\a +\b )}\,.
\ee

We are now ready to define a canonical basis in ${\cal V}_{\bf p}\,.$
It consists of two pieces,
\ba
&&q^{kl} E^{(m)}_1E^{(k)}_2E^{(\ell )}_1\,|{\bf p}; 0 , 0 , 0 {\cal i}\,
:=
\, |{\bf p}; m, k, \ell{\cal i}^{(1)}\,,\nonumber\\
&&q^{kl} E^{(\ell)}_2E^{(k)}_1E^{(m)}_2\,|{\bf p}; 0 , 0 , 0 {\cal i}\,
:=
\, |{\bf p}; \ell , k, m {\cal i}^{(2)}
\lb{two}
\ea
( $0\le \ell , m\,,\ \ell +m\le k\le h-1\,$) related by
\be
\lb{relat12}
|{\bf p}; m, \ell +m, \ell {\cal i}^{(1)} \, =\,
|{\bf p}; \ell , \ell + m, m {\cal i}^{(2)}\,.
\ee
Using (\ref{l1l2}) and (\ref{defVbfp}) we find
\ba
\lb{canH3}
&&{(q^{H_1}\, -\, q^{2m+2\ell -k+1-p_{23}})\,
|{\bf p}; m, k, \ell {\cal i}^{(1)} = 0}\,,\nonumber \\
&&{(q^{H_1}\,-\, q^{2k-m-\ell +1-p_{23}})\,
|{\bf p}; \ell , k, m {\cal i}^{(2)} = 0}\,,\nonumber\\
&&{(q^{H_2}\, -\, q^{2m+2\ell -k+1-p_{12}})\,
|{\bf p}; \ell , k, m {\cal i}^{(2)} = 0}\,,\nonumber \\
&&{(q^{H_2}\, -\, q^{2k-m-\ell +1-p_{12}})\,
|{\bf p}; m, k, \ell {\cal i}^{(1)} = 0}\,;
\ea
\ba
&&E_1\, |{\bf p}; m, k, \ell {\cal i}^{(1)}
= (m+1)_+\, |{\bf p}; m+1, k, \ell {\cal i}^{(1)}
\quad {\rm for}\, k>m+\ell\,,\nonumber\\
&&\nonumber\\
&&E_1\, |{\bf p}; \ell , k, m {\cal i}^{(2)}
= (k-\ell +1)_+ |{\bf p}; \ell , k+1, m {\cal i}^{(2)} +
\nonumber \\
&&+q^{2(k-\ell +1)}
(1-\d_{\ell 0}) (m+1)_+
|{\bf p}; \ell -1, k+1, m+1 {\cal i}^{(2)}
\quad{\rm for}\, k\ge m+\ell\,,\nonumber\\
&&\nonumber\\
&&E_2\, |{\bf p}; m, k, \ell {\cal i}^{(1)}
= q^{m-\ell} (k-m+1)_+ |{\bf p}; m, k+1, \ell {\cal i}^{(1)} +
\nonumber\\
&&+ \bq^{m+\ell} (1-\d_{m 0})
(\ell +1)_+ |{\bf p}; m-1, k+1, \ell +1
{\cal i}^{(1)}
\quad{\rm for}\, k\ge m+\ell\,,\nonumber\\
&&\nonumber\\
&&E_2\, |{\bf p}; \ell , k, m {\cal i}^{(2)} =
\bq^k (\ell +1)_+ |{\bf p}; \ell +1, k, m {\cal i}^{(2)}
\quad {\rm for}\, k>m+\ell\,;
\lb{canE3}\\
&&\phantom{.}\nonumber\\
&&F_1\, |{\bf p}; m, k, \ell {\cal i}^{(1)} =\nonumber\\
&&= q^{2-p_{23}-k+2\ell}(1-\d_{m 0})
(p_{23}-m+k-2\ell )_+
|{\bf p}; m-1, k, \ell {\cal i}^{(1)} + \nonumber\\
&&+q^{2-p_{23}+k}(1-\d_{\ell 0})
(p_{23}-\ell )_+
|{\bf p}; m, k, \ell -1 {\cal i}^{(1)}
\quad {\rm for}\, k\ge m+\ell\,,\nonumber\\
&&\phantom{.}\nonumber\\
&&F_1\, |{\bf p}; \ell , k, m {\cal i}^{(2)}
=\nonumber\\
&&=q^{2-p_{23}+\ell -m}(p_{23}-k+m)_+
|{\bf p}; \ell , k-1, m {\cal i}^{(2)}
\quad{\rm for}\, k>m+\ell\,,\nonumber\\
\nonumber\\
&&F_2\, |{\bf p}; \ell , k, m {\cal i}^{(2)}
=\nonumber\\
&&= q^{2-p_{12}+2m}(1-\d_{\ell 0})
(p_{12}-\ell +k-2m)_+
|{\bf p}; \ell -1, k, m {\cal i}^{(2)}
+ \nonumber\\
&&+q^{2-p_{12}}(1-\d_{m 0}) (p_{12}-m)_+
|{\bf p}; \ell , k, m-1{\cal i}^{(2)}
\quad{\rm for}\, k\ge m+\ell\,,\nonumber\\ \nonumber\\
&&F_2\, |{\bf p}; m, k, \ell {\cal i}^{(1)}
=\nonumber\\
&&= q^{2-p_{12}}(p_{12}-k+\ell )_+
|{\bf p}; m, k-1, \ell {\cal i}^{(1)}
\quad{\rm for}\, k>m+\ell\,.
\lb{canF3}
\ea
Details of the calculations can be found in the Appendix.

The dimension $d_3(h)\,$ (\ref{d3hPBW}) can be recovered by a canonical
basis computation too; we have
\be
\lb{d3h}
d_3 (h) = 2 \sum_{\l_1 = 0}^{h-1} \sum_{\l_2 = 0}^{\l_1-1}
(\l_2 +1 )
+ \sum_{\l = 0}^{h-1} (\l +1 )
= \sum_{\ell = 1}^{h}\, \ell^2 = \frac{h(h+1)(2h+1)}{6}\,.
\ee

\vspace{5mm}

\noindent
{\bf Remark 5.1~} Consider the algebra $U_{(h)}^E\,$ obtained from $U_q^E\,$
by
factoring the latter with respect to the relations $E_a^h = 0 = [ h H ]\,$ and adding
the new elements $E_a^{(h)}\,$ (of \cite{L}) such that
\be
\lb{newel}
[q^{H_a} , E_a^{(h)} ] = 0\,,\quad [E_a , E_a^{(h)} ] = 0\,,\quad [ q^{H_{3-a}} ,
E_a^{(h)} ]_+ = 0\,,\quad a=1,2\,.
\ee
We can define a periodic extension of ${\cal V}_{\bf p}\,$ if we impose
the relations
\be
\lb{perF}
[E_{3-a} , E_a^{(h)} ]_+ {\cal V}_{\bf p} = 0\,,\
[F_a, E_b^{(h)} ] {\cal V}_{\bf p} = 0\,,\
(E_2^{(h)} E_1^{(h)} - (-1)^h E_1^{(h)} E_2^{(h)}) {\cal V}_{\bf p} = 0
\ee
and the periodicity condition
\be
\lb{percond}
\left( \left( E_a^{(h)}\right)^2 - 1 \right) {\cal V}_{\bf p} = 0\,.
\ee
(In view of (\ref{4.6}), $( E_a^{(h)} )^2 = 2 E_a^{(2h)}\,.$) Indeed,
there is an ideal ${\cal J}_{(h)}\,$ of $U^E_{(h)}\,$ generated by
$$
[E_1 , E_2^{(h)} ]_+\,,\quad
[E_2 , E_1^{(h)} ]_+\,,\quad
E_2^{(h)} E_1^{(h)} - (-1)^h E_1^{(h)} E_2^{(h)}\,.
$$
Then Eqs. (\ref{PBW-can}), (\ref{X}) imply that, in
$U^E_{(h)} / {\cal J}_{(h)}\,,$
\ba
&&E_1^{(h-\b )} E_{21}^{[\b ]} =
\sum_{\s = 0}^\b q^{\s (\s -1)} E_2^{(\b -\s )}
E_1^{(h)} E_2^{(\s )} =\nonumber \\
&&= \sum_{\s = 0}^\b (-1)^\s q^{\s (\s -1)} {\b\choose\s}_+ E_2^{(\b )}
E_1^{(h)} = 0
\lb{rel}
\ea
for any $\b >0\,,$ where we are using the relation
$E_1^{(h)} E_2^{(\s )} = (-1)^\s E_2^{(\s )} E_1^{(h)}\,$
and the identity
\be
\lb{q-ver}
\sum_{\s = 0}^\b (-1)^\s q^{\s (\s -1)} {\b\choose\s}_+
= \d_{\b 0}\quad {\rm for}\
\b\in \Z_+
\ee
(essentially, a $q$-version of $(1-1)^\b = 0\,$ for $\b > 0\,$). It follows, in particular, that
\be
\lb{E21h}
E_{21}^{[h]} |{\bf p}; 0,0,0{\cal i} = 0\,.
\ee

\vspace{5mm}

\noindent
{\bf Remark 5.2~} The space ${\cal V}_{\bf p}\,$ can be defined -- in
terms of the
PBW basis -- for $U_q(s\ell_n )\,$ with $n>3\,$ as well. The basis involves
${n\choose 2}\,$ exponents $\a_{ij} = \a_{ji} \in \Z_+\,$
satisfying $\sum_j\, \a_{ij} \,\le h-1\,.$ The
dimension $d_n(h)\,$ of ${\cal V}_{\bf p}\,$ is a polynomial in $h\,$ of
degree
${n\choose 2}\,;$ for $n=4\,$ it is given by
\be
\lb{d4h}
d_4(h) = \frac{1}{6!}\, h (h+1)^2 (h+2) (h+3) (11h+4)\,.
\ee

\vspace{5mm}

\subsection{BRS intertwiners, singular vectors and invariant subspaces}

\medskip

The presence of two (equivalent) bases in ${\cal V}_{\bf p}\,$ is an
asset:
it enables us to use for each problem the one which is better adapted to
its solution. We shall illustrate this fact by writing down the BRS
intertwiners
\be
\lb{inter3}
Q_{\bf p}:\, {\cal V}_{w_L \bf p}\ \rightarrow \ {\cal V}_{\bf
p}\,,\qquad Q_{{\bf h}+w_L \bf p}:\, {\cal V}_{\bf p}\ \rightarrow \
{\cal V}_{{\bf h}+w_L \bf p}
\ee
in the PBW basis and the singular vectors in ${\cal V}_{\bf p}\,$ in the
canonical basis.

Let us start with $Q_{\bf p}\,.$ From $q^{H_i}\,$ invariance
(\ref{perCart}) it follows that
\be
\lb{Qbasic}
Q_{\bf p} |\;w_L {\bf p}\;;\a,\b,\g{\cal i} = \sum_\rho\; f_\rho
(\a,\b,\g;{\bf p} ) |\; {\bf p}\;; \a+\rho,\b -\rho + p_{13}, \g
+\rho{\cal i}\,.
\ee
The $E_1\,$ and $E_{21}\,$ invariance (cf. (\ref{E3})) implies
\be
\lb{E1inv}
f_\rho (\a,\b,\g;{\bf p} ) = q^{(\a+\b )\rho}
\left[\matrix{\b+p_{13}-\rho\cr \b\cr}\right]
\left[\matrix{\a+\rho\cr \a\cr}\right] f_\rho (0,0,\g;{\bf p}
)\,.
\ee
To find $f_\rho (0,0,\g;{\bf p} ) \equiv f_\rho (\g ;{\bf p})\,,$
it is convenient to use the equations following from $E_2\,$ and
$F_1\,$ invariance since they contain the same triples of various
$\rho\,$ and $\g\,$ combinations). Applying $E_2\,,$ one gets
\ba
\lb{E2inv}
&&q^\b \;(\g +1)_+ \;f_\rho (\a,\b,\g +1;{\bf p} )
- q^{-\b}\; (\b+1)_+\; f_{\rho +1}(\a -1,\b +1,\g;{\bf p} ) =
\nonumber\\
&&= q^{\b -2\rho +p_{13}} (\g +\rho +1)_+ f_\rho (\a,\b,\g;{\bf p}
) -\\ &&-\; q^{-\b -p_{13}}\; (\b -\rho +p_{13})_+\;
f_{\rho +1}\;(\a,\b,\g ;{\bf p})\,,
\nonumber
\ea
whereas $F_1\,$ invariance implies
\ba
\lb{F1inv}
&&q^{-\a-\b-p_{12}+1} \;[\g -\a-\b-p_{12}]\; f_{\rho +1}(\a-1,\b,\g
];{\bf p} ) =\nonumber\\ &&= q^{-\rho -\a-\b -p_{12}}\; [\g +\rho
-\a-\b
-p_{12}+1]\; f_{\rho+1} (\a,\b,\g;{\bf p} )+\\
&&+\; q^{-\rho}\; [\g +\rho +1 ] \; f_\rho (\a,\b,\g ;{\bf p})-
[\g+1]\; f_{\rho}(\a,\b -1,\g+1;{\bf p} )\,.\nonumber
\ea
After some algebra one obtains the following simple and
selfconsistent -- e.g. not containing $\a\,$ and $\b\,$ --
recurrence relations for the function $f_\rho (\g ; {\bf p})\,$:
\ba
\lb{frhogamma}
f_{\rho +1} (\g ;{\bf p})\, &=&\, - \;{{(\g +\rho +1)_+}\over{(\g
+\rho
-p_{12} +1)_+}}\; f_\rho (\g ;{\bf p})\,,\nonumber\\ \\
f_{\rho} (\g +1;{\bf p})\, &=&\, {{[\g +\rho +1][\g +p_{23}
+1]}\over{[\g +1][\g +\rho
-p_{12} +1]}}\;  f_\rho (\g ;{\bf p})\,.\nonumber
\ea
Solving the recurrence relations (\ref{frhogamma}) for $f_\rho (\g
;{\bf p})\,$ and putting the latter in (\ref{E1inv}), one can bring
the final result for $f_\rho (\a,\b,\g;{\bf p} )\,$ to the form
\ba
\lb{fininv}
&&f_\rho (\a,\b,\g; {\bf p} ) =\\
&&= (-q^{(\a+\b +p_{12})})^\rho
\left[\matrix{\b+p_{13}-\rho\cr \b\cr}\right]
\left[\matrix{\a+\rho\cr \a\cr}\right]
\left[\matrix{\g+\rho\cr p_{12}\cr}\right]
\left[\matrix{\g+p_{23}\cr p_{23}\cr}\right]\;
{f ({\bf p})} = \nonumber\\
&&= (-q^{(\a+\b +p_{12})})^\rho\; {{[\a+\rho]!
[\b +p_{13}-\rho]![\g+\rho]![\g+p_{23}]!}\over{[\rho]![\a]![\b]![\g]!
[p_{13}-\rho]![\g-p_{12}+\rho]!}}\; \frac{f({\bf
p})}{[p_{12}]![p_{23}]!}\,.\nonumber
\ea
One sees that the effective range of summation over $\rho\,$ in
(\ref{Qbasic}) (for $0 < p_{13} < h\,$) is from
${\rm max}\; (p_{12}-\g , 0)\,$ to ${\rm min}\; (p_{13} , h-\a , h-\g
)\,$.

One can find quite analogously the corresponding expression for the
action $Q_{{\bf h}+w_L {\bf p}}\,.$ It turns out that
\ba
\lb{Qwbasic}
&&Q_{{\bf h}+w_L \bf p} |\;{\bf p}\;;\a,\b,\g{\cal i} = \sum_\rho\; f_\rho
(\a,\b,\g; h-p_{23},h- p_{12} )\times\nonumber\\
&& \times|\;
w_L{\bf p}\;; \a+\rho,\b +2h
- p_{13} - \rho , \g +\rho{\cal i}\,.
\ea
It is easy to see that the successive application of $Q_{\bf p}\,$
and $Q_{{\bf h}+w_L \bf p}\,$ gives zero, i.e.
\be
\lb{n3Q2}
Q_{\bf p} Q_{{\bf h}+w_L \bf p} \; = \; 0 \;=\; Q_{{\bf h}+w_L \bf p}
Q_{\bf p}\,.
\ee
This follows from the explicit form of the coefficient functions
$f_\rho (\a,\b,\g; {\bf p} )\,.$ Indeed (for $f({\bf p}) = 1\,$)
\ba
\lb{QwQ}
&& Q_{{\bf h}+w_L \bf p} Q_{\bf p}
|\, {\bf h}+w_L {\bf p}\;;\a,\b,\g{\cal i}
= \sum_\sigma\;|\; w_L {\bf p}\;;
\a+\sigma,\b + 2h -\sigma, \g +\sigma {\cal i}\times\nonumber\\
&&\times \sum_\rho\; f_\rho (\a,\b,\g;{\bf p} ) f_{\sigma-\rho}
(\a+\rho,\b+p_{13}-\rho,\g+\rho;h-p_{23},h-p_{12})\nonumber\\
\ea
and
\ba
\lb{ff}
&&f_\rho (\a,\b,\g;{\bf p} )\;
f_{\sigma-\rho}(\a+\rho,\b+p_{13}-\rho,\g+\rho;h-p_{23},h-p_{12})\;=
\nonumber\\
&&={{[\a+\sigma]![\b+2h-\sigma]![\g+\sigma]!
[\g+\rho+h-p_{12}]!}\over{[\rho]![\a]![\b]![\g]!
[\g+\rho-p_{12}]![\sigma-\rho]!}[p_{12}]![p_{23}]!
[h-p_{12}]![h-p_{23}]!}\times\nonumber\\ &&\times {{[\g+p_{23}]!
}\over { [p_{13}-\rho]! [2h-p_{13}+\rho-\sigma]!
[\g+\sigma-h+p_{23}]!} } \equiv 0
\ea
(due to factorials of integers greater or equal to $h\,$ in the
numerator, or factorials of negative integers in the denominator).

For similar reasons
\be
\lb{QQw}
Q_{\bf p} Q_{{\bf h}+w_L \bf p} |\;{\bf p}\;;\a,\b +p_{13},\g{\cal
i}
= \sum_\sigma\;|\; {\bf h}+ {\bf p}\;;
\a+\sigma,\b + p_{13}
-\sigma, \g +\sigma {\cal i}\times
\ee
$$\times \sum_\mu\; f_\mu (\a,\b+p_{13},\g; h-p_{23}, h-p_{12})
f_{\sigma-\mu} (\a+\mu,\b+2h-\mu,\g+\mu; {\bf p})
$$
vanishes identically, too.

One should expect the vector $ Q_{\bf p}\;|\;w_L {\bf p}; 0,0,
0{\cal i} \in {\cal V}_{\bf p}\,$ to be {\em singular}.
In view of (\ref{Qbasic}) and (\ref{fininv}), it can be shown
to be proportional to
\be
\lb{s}
|s_{\bf p}{\cal i} = \sum_{\a=p_{12}}^{p_{13}} (-1)^\a
\left(\matrix{\a\cr p_{12}\cr} \right)_+
|{\bf p}; \a,p_{13}-\a, \a {\cal i}\,.
\ee
The dual formulas are
\ba
&&|s_{{\bf h}+w_L\bf p}{\cal i} = \sum_{\a=h-p_{23}}^{2h-p_{13}} (-1)^\a
\left(\matrix{\a\cr h-p_{23}\cr}\right)_+\,
|{\bf h}+w_L {\bf p}; \a,2h-p_{13}-\a, \a{\cal i}\,\sim \nonumber\\
&&\sim\, Q_{{\bf h}+w_L\bf p}|{\bf p}; 0,0,0 {\cal i}.
\lb{sw}
\ea
Applying (\ref{F3}), one can check that the weight vectors
(\ref{s}) and (\ref{sw}) satisfy indeed
\be
\lb{sing}
F_a\; |s_{\bf p}{\cal i}\; =\; 0\; =\; F_a\; |s_{{\bf h}+w_L\bf p}{\cal
i}\,,\ \quad a=1,2\,.
\ee
The expressions for the singular vectors
become particularly elegant in the canonical basis. Noting that for
$m+\ell = k\,$ Eqs. (\ref{can-PBW}) reduce to
\be
\lb{can-PBW1}
q^{k\ell}
E_1^{(k-\ell)}E_2^{(k)}E_1^{(\ell )}
\equiv
q^{k\ell}
E_2^{(\ell )} E_1^{(k)} E_2^{(k-\ell )}
=
\sum_{j=0}^{\ell}
(-1)^j {{k-j}\choose{m}}_+ E_1^{(k
-j)}E_{21}^{[j]}E_2^{(k-j)}
\ee
and substituting $m=p_{12}\,,\ \ell = p_{23}\,,\ k=p_{13}\,,
\ j=p_{13}-\a\,,$ one obtains
\ba
\lb{can-PBW2}
&&q^{p_{13}p_{23}}\, E_1^{(p_{12})}E_2^{(p_{13})}E_1^{(p_{23})}
\, \equiv \,
q^{p_{13}p_{23}}\, E_2^{(p_{23})} E_1^{(p_{13})}
E_2^{(p_{12})}\, =\nonumber\\
&&= (- 1)^{p_{13}} \sum_{\a=p_{12}}^{p_{13}}
(-1)^{\a} {{\a}\choose{p_{12}}}_+
E_1^{(\a)}E_{21}^{[p_{13}-\a]}E_2^{(\a)} \,.
\ea
Hence, for $|s_{\bf p}{\cal i} \in {\cal V}_{\bf p}\,$ we have
\be
\lb{s1}
|{\bf p}; p_{12}, p_{13}, p_{23} {\cal i}^{(1)}\,\equiv\,
|{\bf p}; p_{23}, p_{13}, p_{12} {\cal i}^{(2)}\,
=\, (- 1)^{p_{13}} |s_{\bf p}{\cal i}\,.
\ee

The identification of invariant subspaces and quotients of
${\cal V}_{\bf p}\,$ is facilitated by the knowledge of the invariant
hermitean form (which majorizes the invariant bilinear form -- cf. Remark
3.4). Noting that $(E_a^{(m)})^* = \frac{1}{(m)_-!} F_a^m\,$ and observing
the identity $(m)_+ (m)_- = [m]^2\,,$ we can write the following
expression for the norm square of canonical basis vectors:
\be
\lb{norm-can}
\| |{\bf p}; m,k,\ell{\cal i}^1 \|^2 =
{\cal h}{\bf p};0,0,0 | F_1^{[\ell ]} F_2^{[k]} F_1^{[m]}
E_1^{[m]} E_2^{[k]} E_1^{[\ell ]} |{\bf p}; 0,0,0{\cal i}
\ee
and a similar expression for $1\ \leftrightarrow\ 2\,$ and $\ell\
\leftrightarrow\ m\,$ which is completely determined if we set ${\cal h}
{\bf p};0,0,0 | {\bf p}; 0,0,0 {\cal i} = 1\,$ (for $1<p_{13}<h\,$).

\vspace{5mm}

\noindent
{\bf Remark 5.3}~ We surely can also compute the invariant bilinear and
hermitean forms in the PBW basis. For instance, one can derive the
relation
\be
\lb{norm-bil}
{\cal h}{\bf p}; \a ,\b ,\g |\,{\bf p}; \a ,\b ,\g {\cal i}=
\sum_{\sigma = {\rm max}(\alpha -\beta ,0)}^\alpha (-1)^{\alpha
+\sigma}\,\times
\ee
$$\times
\left[\matrix{ \alpha +\gamma -\sigma \cr \gamma \cr }\right]^2
\left[\matrix{ p_{23}+\gamma -\beta -1\cr \sigma \cr }\right]
\left[\matrix{ p_{13}+\sigma -\alpha -\gamma -2\cr \beta -\alpha +\sigma \cr }
\right]
\left[\matrix{ p_{12}-1\cr \alpha +\gamma-\sigma \cr }\right] \,.
$$
Here the advantage of the canonical basis becomes manifest: for a singular
vector of the type (\ref{s1}) the vanishing of its norm square
(\ref{norm-can}) is an immediate consequence of (\ref{sing}). By contrast,
the (bilinear) squares (\ref{norm-bil}) of PBW basis vectors entering the
expansion (\ref{s}) are not, in general, zero -- only the resulting sum of
inner products shoud vanish.

\vspace{5mm}

A calculation similar to the (canonical basis) computation of $d_3(h)\,$
yields the dimension of the image of $Q_{\bf p}\,$ in ${\cal V}_{\bf
p}\,:$
\be
\lb{dimim}
{\rm dim}\ ({\rm Im}\, Q_{\bf p} )
= {\rm dim}\, \left( {\rm Ker}\, Q_{{\bf h}+w_L{\bf p}}\,\right)
=\frac{1}{6} (h-p_{13}) (h-p_{13}+1)
(2h-2p_{13}+1)\,.
\ee
This invariant subspace lies in the kernel of the hermitean
form on ${\cal V}_{\bf p}\,$ but does not exhaust it.

A systematic study of the structure of invariant subspaces and
subquotients of ${\cal V}_{\bf p}\,$
is left for future work.

\vspace{5mm}

\noindent
\section*{Acknowledgements}
A major part of the work has been completed while all three authors
were in Trieste. P.F. acknowledges the support of the
University of Trieste and the Italian Ministry of
University, Scientific Research and Technology (MURST).
L.H. thanks DFT, Universit\`a di Trieste and
INFN, Sezione di Trieste, for their hospitality and support.
I.T. acknowledges the hospitality of SISSA.
L.H. and I.T. have been supported in part by the Bulgarian
National Foundation for Scientific Research under contract F-828 and
enjoyed the hospitality of the Theory Division of CERN during the final
stage of this work.

\vspace{5mm}
\noindent
\section*{Appendix}
\setcounter{equation}{0}
\renewcommand\theequation{A.\arabic{equation}}

Here we give the details of the most involved calculation in the
transition from PBW to canonical basis that contains all the needed
technical steps. It is assumed that $k\ge m+\ell.$
$$E_2 \;E_1^{(m)}E_2^{(k)}E_1^{(\ell )} |{\bf p}; 0,0,0{\cal i}=$$
$$= {\bar q}^{k\ell}
\sum_{j=0}^{\ell}(-1)^j\left({m+\ell -j\atop m}\right)_+
E_2 |{\bf p};m+\ell -j,j,k-j {\cal i} =$$
$$={\bar q}^{m+\ell (k+1)}
\sum_{j=0}^{\ell}(-1)^j\left({m+\ell -j\atop m}\right)_+\times$$
$$\times \{\, q^{2j}(k-j+1)_+|{\bf p};m+\ell -j,j,k-j+1 {\cal i} - $$
$$ - (j+1)_+|{\bf p};m+\ell -j-1,j+1,k-j {\cal i}\,\} =$$
$$= {\bar q}^{m+\ell (k+1)}
\sum_{j=0}^{\ell +1}(-1)^j |{\bf p};m+\ell -j,j,k-j+1 {\cal i} \times$$
$$\times  \left\{\left({m+\ell -j\atop m}\right)_+
q^{2j}(k-j+1)_+ + \left({m+\ell +1-j\atop m}\right)_+(j)_+\right\}
= $$
$$ = {\bar q}^{m+\ell (k+1)}
\sum_{j=0}^{\ell +1} \,(-1)^j
\times$$
$$
\times
\left\{\left({m+\ell -j\atop
m}\right)_+
q^{2j}(k-j+1)_+ + \left({m+\ell +1-j\atop m}\right)_+(j)_+\right\}\times$$
$$\times \sum_{n=0}^j (-1)^{j-n}\left({m+\ell -j+n\atop n}\right)_+
\times$$
$$\times
q^{n(n-1)+(j-n)(k+1)}E_1^{(m+\ell -j+n)} E_2^{(k+1)} E_1^{(j-n)}
|{\bf p}; 0,0,0 {\cal i} =$$
(now we change the summation index $n$ by $j-n$)
$$ = {\bar q}^{m+\ell (k+1)}
\sum_{j=0}^{\ell +1}(-1)^j
\times$$
$$\times
\left\{\left({m+\ell -j\atop m}\right)_+
q^{2j}(k-j+1)_+ + \left({m+\ell +1-j\atop m}\right)_+(j)_+\right\}\times$$
$$\times \sum_{n=0}^j (-1)^n\left({m+\ell -n\atop {j-n}}\right)_+
q^{(j-n)(j-n-1)+n(k+1)}\times$$
$$\times\, E_1^{(m+\ell -n)} E_2^{(k+1)} E_1^{(n)}
|{\bf p}; 0,0,0 {\cal i} =$$
(here we can change the upper summation limit in the sum over $n$
from $j$ to $\ell +1$ since the binomial coefficient
$\left({m+\ell -n\atop {j-n}}\right)_+$
is automatically zero when $n\ge j$; after that we
can exchange the orders of summation over $n$ and $j$)
$$ = {\bar q}^{m+\ell (k+1)}
\sum_{n=0}^{\ell +1}(-1)^n q^{n(k+1)}\; \sum_{j=0}^{\ell +1} (-1)^j
\left({m+\ell -n\atop {j-n}} \right)_+ q^{(j-n)(j-n-1)} \times$$
$$ \times \left\{ \left({{m+\ell -j}\atop m}\right)_+ q^{2j} (k-j+1)_+
+\left({m+\ell +1-j\atop m}\right)_+(j)_+\right\}\times$$
\be
\lb{A.1}
\times E_1^{(m+\ell -n)} E_2^{(k+1)} E_1^{(n)} |{\bf p}; 0,0,0 {\cal
i} \,.
\ee
Let us now compute the sum over $j$:
$$\sum_{j=0}^{\ell +1} (-1)^j
\left( {m+\ell -n\atop {j-n}} \right)_+ q^{(j-n)(j-n-1)} \times$$
$$\times \left\{ \left({{m+\ell -j}\atop m}\right)_+ q^{2j} (k-j+1)_+
+\left({m+\ell +1-j\atop m}\right)_+(j)_+\right\}=$$
$$=\sum_{j=0}^{\ell +1} (-1)^j q^{(j-n)(j-n-1)}{{(m+\ell -n)_+!}\over
{(m)_+!(j-n)_+!(\ell +1-j)_+!}}\times$$
$$\times\left\{ q^{2j}(k+1-j)_+(\ell +1-j)_+ + (j)_+(m+\ell +1-j)_+\right\}=$$
(crucial observation:
$$q^{2j}(k+1-j)_+(\ell +1-j)_+ + (j)_+(m+\ell
+1-j)_+\, =$$
$$ = (m+\ell
+1-j)_+(k-m+1)_++q^{2(k-m+1)}(m)_+(\ell-k+m)_+\,,$$ the second
expression being ``$q$-linear" in $j\,$)
$$={{(m+\ell -n)_+!}\over{(m)_+!(\ell +1-n)_+!}}
\sum_{j=0}^{\ell +1} (-1)^j q^{(j-n)(j-n-1)}
\left({\ell +1-n\atop{j-n}}\right)_+\times$$
$$\times\left\{(m+\ell +1-j)_+(k-m+1)_++q^{2(k-m+1)}(m)_+(\ell -k+m)_+
\right\}=$$
(denoting $j-n$ by $j$)
$$={{(m+\ell -n)_+!}\over{(m)_+!(\ell +1-n)_+!}}(-1)^n
\sum_{j=0}^{\ell +1-n} (-1)^j q^{j(j-1)}
\left({\ell +1-n\atop{j}}\right)_+\times$$
$$\times\left\{(m+\ell -n+1-j)_+(k-m+1)_++q^{2(k-m+1)}(m)_+(\ell -k+m)_+
\right\}=$$
(using
\be
\lb{A.2}
\sum_{j=0}^n (-1)^j q^{j(j-1)} {n\choose j}_+ (M-j)_+ =
(M)_+\delta_{n 0} + q^{2(M-1)} \delta_{n 1}\,
\ee
and (\ref{q-ver}) with $\b\,\to\, n\,,\ \s\,\to\, j\,$)
$$={{(m+\ell -n)_+!}\over{(m)_+!(\ell +1-n)_+!}}(-1)^n
\{ q^{2m}(k-m+1)_+\delta_{\ell n}+$$
$$+(m)_+\left[ (k-m+1)_+ +q^{2(k-m+1)}(\ell
-k+m)_+\right]\delta_{\ell +1\; n}\}=$$
$$={{(m+\ell -n)_+!}\over{(m)_+!(\ell +1-n)_+!}}(-1)^n\left\{
q^{2m}(k-m+1)_+\delta_{\ell n} +
(m)_+(\ell +1)_+\delta_{\ell +1\, n}\right\}=$$
\be
\lb{A.3}
=(-1)^n \left\{
q^{2m} (k-m+1)_+\delta_{\ell n}+ (\ell +1)_+\delta_{\ell +1\; n}
\right\}\,.
\ee
Combining (\ref{A.1}) and (\ref{A.3}), we get eventually (for $k\ge m+\ell$)
\be
\lb{A.4}
E_2 E_1^{(m)}E_2^{(k)}E_1^{(\ell )} |{\bf p}; 0,0,0 {\cal i} =
\sum_{n=0}^{\ell +1}{\bar q}^{(\ell -n)(k+1)+m}\times
\ee
$$\times\left\{ q^{2m}
(k-m+1)_+\delta_{\ell n}+(\ell +1)_+\delta_{\ell +1\; n}\right\}
E_1^{(m+\ell -n)} E_2^{(k+1)} E_1^{(n)} |{\bf p}; 0,0,0 {\cal i} =$$
$$=q^m (k-m+1)_+ E_1^{(m)}E_2^{(k+1)}E_1^{(\ell )} |{\bf p}; 0,0,0 {\cal
i} \, +$$
$$+\, q^{k-m+1}(\ell +1)_+ E_1^{(m-1)}E_2^{(k+1)}E_1^{(\ell +1)}
|{\bf p}; 0,0,0 {\cal i}\,.$$


\vspace{5mm}

\begin{thebibliography}{000}

\bibitem{W}
E. Witten,
Commun. Math. Phys. {\bf 92} (1984) 455-472.

\bibitem{KZ}
V.G. Knizhnik and A.B. Zamolodchikov,
Nucl. Phys. {\bf B247} (1984) 83-103;\\
I.T. Todorov, International Conference on Differential Geometric
Methods in Physics, Shumen, 1984, Eds. H.-D. Doebner and T.
Palev (World Scientific, Singapore) pp. 297-347; Phys. Lett. {\bf
B153} (1985) 77-81.

\bibitem{B}
O. Babelon, Phys. Lett. {\bf B215} (1988) 523-529;\\
B. Blok, Phys. Lett. {\bf B233} (1989) 359-362.

\bibitem{F1}
L.D. Faddeev, Commun. Math. Phys. {\bf 132} (1990) 131-138;\\
A. Alekseev and S. Shatashvili, Commun. Math. Phys. {\bf 133}
(1990) 353-368.

\bibitem{F2}
L.D. Faddeev, {\em Quantum symmetry in conformal field theory by
Hamiltonian methods}, Carg\`ese lectures 1991, {\bf in:}
New Symmetry Principles in Quantum Field
Theory, Ed. J. Fr\"ohlich et al., Plenum Press, NY (1992),
pp. 159--175;\\
A.Yu. Alekseev, L.D. Faddeev and M.A. Semenov--Tian--Shansky,
Commun. Math. Phys. {\bf 149} (1992) 335-345.

\bibitem{G}
K. Gaw\c{e}dzki, Commun. Math. Phys. {\bf 139} (1991) 201-213.

\bibitem{FG}
F. Falceto and K. Gaw\c{e}dzki, J. Geom. Phys. {\bf 11} (1993)
251-279.

\bibitem{FHT1}
P. Furlan, L.K. Hadjiivanov and I.T. Todorov,
{\em Canonical approach to the quantum WZNW model},
ICTP Trieste and ESI Vienna preprint IC/95/74, ESI 234 (1995).

\bibitem{FHT2}
P. Furlan, L.K. Hadjiivanov and I.T. Todorov,
Nucl. Phys. {\bf B474} (1996) 497-511, {\tt hep-th/9602101}.

\bibitem{FHT3}
P. Furlan, L.K. Hadjiivanov and I.T. Todorov,
Int. J. Mod. Phys. {\bf A12} (1997) 23-32, {\tt hep-th/9610202}.

\bibitem{FHIOPT}
P. Furlan, L.K. Hadjiivanov, A.P. Isaev, O.V. Ogievetsky,
P.N. Pyatov and I.T. Todorov,
{\em Quantum matrix algebra for the $SU(n)\,$ WZNW model}, {\tt
hep-th/0003210},
IHES Bures-sur-Yvette preprint IHES/P/00/11, submitted to Commun. Math.
Phys.

\bibitem{Wa}
M. Wakimoto,
Commun. Math. Phys. {\bf 104} (1986) 605-609.

\bibitem{BF}
D. Bernard and G. Felder,
Commun. Math. Phys. {\bf 127} (1990) 145-168.

\bibitem{BMP}
P. Bouwknegt, J. McCarthy and K. Pilch,
Progr. Theor. Phys. Suppl. {\bf 102} (1990) 67-135.

\bibitem{PS90}
V. Pasquier and H. Saleur,
Nucl. Phys. {\bf B330} (1990) 523.


\bibitem{ReshTur}
N. Reshetikhin, V.G. Turaev,
Invent. Math. {\bf 103} (1991) 547-597.

\bibitem{FK}
J. Fr\"ohlich and T. Kerler,
{\em Quantum Groups, Quantum Categories and Quantum Field Theory},
Lecture Notes in Mathematics {\bf 1542}, Springer, Berlin (1993).

\bibitem{FRT}
L.D. Faddeev, N. Yu. Reshetikhin and L.A. Takhtajan,
Algebra i Anal. {\bf 1:1} (1989) 178-206
(English translation: Leningrad Math. J. {\bf 1} (1990) 193-225).

\bibitem{DCK}
C. De Concini and V.G. Kac,
{\em Representations of quantum groups at roots of $1$}, Colloque
Dixmier 1989, pp. 471-506; Progr. in Math. {\bf 92}, Birkh\"auser,
Boston (1990);\\
C. De Concini, V.G. Kac and C. Procesi,
{\em Quantum coadjoint action}, J. Amer. Math. Soc. {\bf 5}
(1992) 151-189.

\bibitem{Ham}
M. Hamermesh,
{\em Group Theory and Its Application to Physical Problems,}
Addison--Wesley, Reading MA (1962).

\bibitem{DV}
M. Dubois-Violette,
K-theory {\bf 14} (1998) 371-404.

\bibitem{DT}
M. Dubois-Violette and I.T. Todorov,
Lett. Math. Phys. {\bf 42} (1997) 183-192, {\tt hep-th/9704069}.

\bibitem{Arnaudon}
D. Arnaudon, Commun. Math. Phys. {\bf 159} (1994) 175-194, {\tt
hep-th/ 9212067}.

\bibitem{L}
G. Lusztig,
Journ. Amer. Math. Soc. {\bf 3} (1990) 447-498;
{\em Introduction to Quantum Groups}, Progress in
Math. {\bf 110}, Birkh\"auser, Boston-Basel-Stuttgart 1993.

\bibitem{K}
M. Kashiwara,
Commun. Math. Phys. {\bf 133} (1990) 249-260.

\bibitem{HIOPT}
L.K. Hadjiivanov, A.P. Isaev, O.V. Ogievetsky,
P.N. Pyatov and I.T. Todorov, {\em Hecke algebraic properties of
dynamical $R$-matrices. Application to related matrix algebras},
J. Math. Phys. {\bf 40} (1999) 427-448, {\tt q-alg/9712026}.

\bibitem{FSoT}
P. Furlan, G.M. Sotkov and I.T. Todorov,
Riv. Nuovo Cim. {\bf 12:6} (1989) 1-202.

\bibitem{SST}
H. Sazdjian, Y.S. Stanev and I.T. Todorov,
J. Math. Phys. {\bf 36} (1995) 2030-2052.

\end{thebibliography}
\end{document}